\documentclass[english,preprint]{elsarticle}
\usepackage[T1]{fontenc}
\usepackage[latin9]{inputenc}
\usepackage{babel}
\usepackage{geometry}
\usepackage{graphicx}
\usepackage{array}
\usepackage{multirow}
\usepackage{float}
\usepackage{rotating}

\makeatletter

\@ifundefined{showcaptionsetup}{}{%
 \PassOptionsToPackage{caption=false}{subfig}}
\usepackage{subfig}
\usepackage{lineno}

\makeatother

\begin{document}

\begin{frontmatter}

\title{Dark Matter Search Backgrounds from Primordial Radionuclide Chain Disequilibrium}

\author[]{D.\,C.~Malling\corref{cor1}}

\author[]{S.~Fiorucci}

\author[]{M.~Pangilinan}

\author[]{J.\,J.~Chapman}

\author[]{C.\,H.~Faham}

\author[]{J.\,R.~Verbus}

\author[]{R.\,J.~Gaitskell}

\cortext[cor1]{Corresponding author: David\_Malling@brown.edu}

\address{Brown University, Department of Physics, 182 Hope Street, Providence, RI 02912, USA}

\begin{abstract}

Dark matter direct-detection searches for weakly interacting massive particles (WIMPs) are commonly limited in sensitivity by neutron and gamma backgrounds from the decay of radioactive isotopes. Several common radioisotopes in detector construction materials are found in long decay chains, notably those headed by $^{238}$U, $^{235}$U, and $^{232}$Th. Gamma radioassay using Ge detectors identifies decay rates of a few of the radioisotopes in each chain, and typically assumes that the chain is in secular equilibrium. If the chains are out of equilibrium, detector background rates can be elevated significantly above expectation. In this work we quantify the increase in neutron and $\gamma$ production rates from an excess of various sub-chains of the $^{238}$U decay chain.

We find that the $^{226}$Ra sub-chain generates $\times$10 higher neutron flux per decay than the $^{238}$U~early sub-chain and $^{210}$Pb sub-chain, in materials with high ($\alpha$,n) neutron yields. Typical gamma screening results limit potential $^{238}$U~early sub-chain activity to $\times$20-60 higher than $^{226}$Ra sub-chain activity. Monte Carlo simulation is used to quantify the contribution of the sub-chains of $^{238}$U to low-energy nuclear recoil (NR) and electron recoil (ER) backgrounds in simplified one~tonne liquid Ar and liquid Xe detectors. NR and ER rates generated by $^{238}$U sub-chains in the Ar and Xe detectors are found after comparable fiducial and multiple-scatter cuts. The Xe detector is found to have $\times$12 higher signal-to-background for 100~GeV WIMPs over neutrons than the Ar detector. ER backgrounds in both detectors are found to increase weakly for excesses of $^{238}$U~early sub-chain and $^{210}$Pb sub-chain relative to $^{226}$Ra sub-chain. Experiments in which backgrounds are NR-dominated are sensitive to undetected excesses of $^{238}$U~early sub-chain and $^{210}$Pb sub-chain concentrations. Experiments with ER-dominated backgrounds are relatively insensitive to these excesses.

\end{abstract}


\end{frontmatter}



\section{Introduction}

\begin{sidewaysfigure}[p]
\begin{centering}
\includegraphics[width=0.75\textwidth]{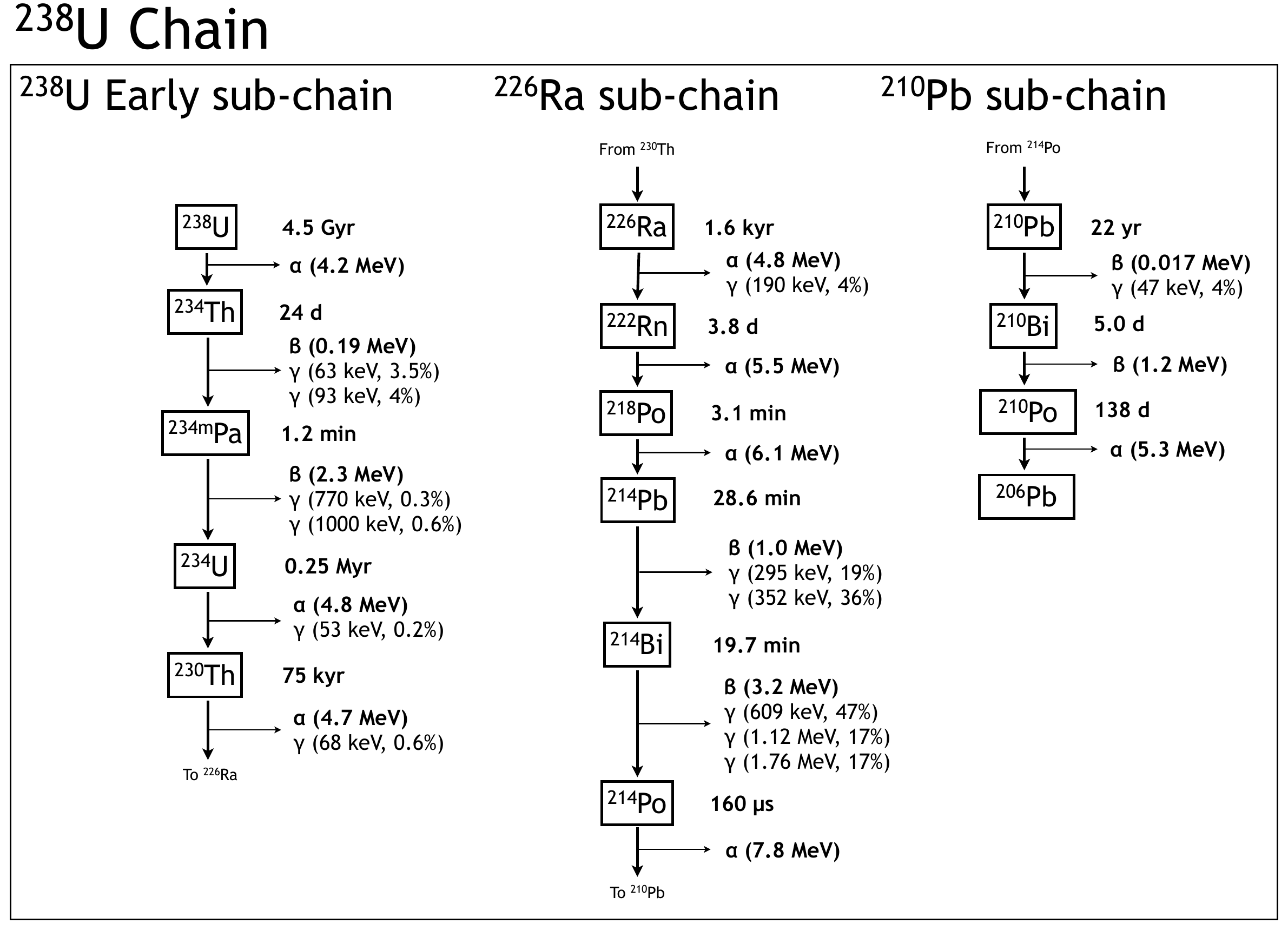}
\par\end{centering}
\caption{\label{238U-decay-chain}The decay chain for $^{238}$U. The decay chain is broken into three sub-chains: the $^{238}$U~early sub-chain, the $^{226}$Ra sub-chain, and the $^{210}$Pb sub-chain. Isotopes are shown with their half-lives and probability for $\alpha$ or $\beta$ emission, if not 100\%. Decays with probability $<$1\% are not shown. Alpha or $\beta$ emission is listed under each isotope, with the mean $\alpha$ energy or $\beta$ decay endpoint given. Energies and intensities are listed for $\gamma$ emission with intensity $>$1\%, with the exception of the $^{238}$U~early chain which lists $\gamma$~rays with intensity $>$0.1\%. Data from \cite{nndc}.}
\end{sidewaysfigure}

\begin{figure}
\begin{centering}
\includegraphics[width=1\textwidth]{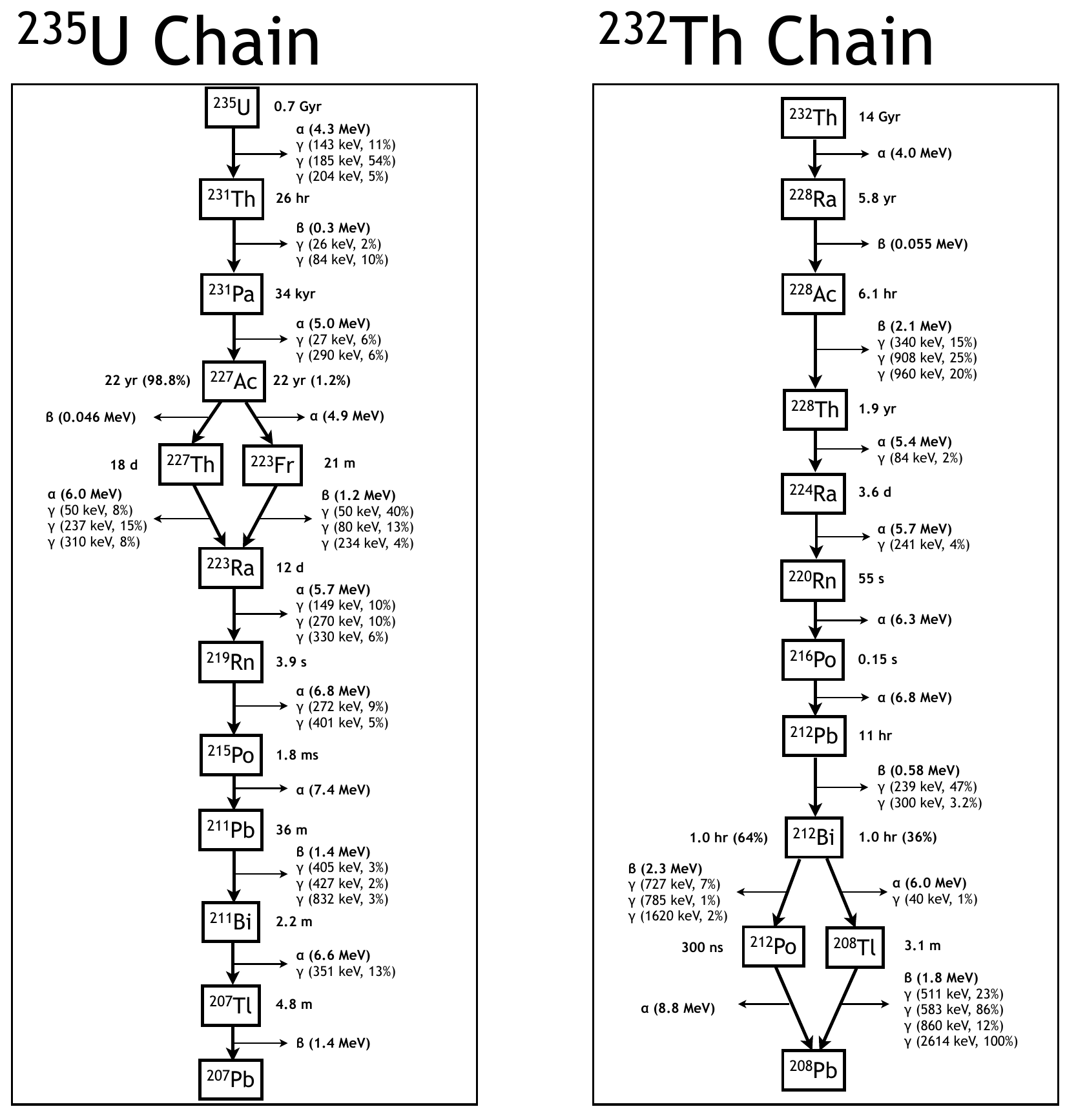}
\par\end{centering}
\caption{\label{235U-232Th-decay-chains}The decay chains for (left) $^{235}$U and (right) $^{232}$Th. Isotopes are shown with their half-lives and probability for $\alpha$ or $\beta$ emission, if not 100\%. Decays with probability $<$1\% are not shown. Alpha or $\beta$ emission is listed under each isotope, with the mean $\alpha$ energy or $\beta$ decay endpoint given. Energies and intensities are listed for $\gamma$ emission with intensity $>$1\%. Data from \cite{nndc}.}
\end{figure}

Direct WIMP dark~matter searches are typically limited by backgrounds caused by neutron and $\gamma$ interactions in the target material. The primary source of these particles is radioactive decay from contaminants in detector materials or the laboratory. The most common of these contaminants which produce notable backgrounds are the primordial radioisotopes $^{238}$U, $^{235}$U and $^{232}$Th. These radioisotopes are the parents of lengthy decay chains, shown in Fig.~\ref{238U-decay-chain} and \ref{235U-232Th-decay-chains}. Decay chain data is taken from \cite{nndc}.

Several isotopes from these chains decay with half-lives much longer than the typical lifetimes of dark~matter experiments ($\gg$10~years). The removal of these long-lived isotopes from the decay chains, e.g. from chemical processing, can lead to quasi-permanent breaks in secular equilibrium. If chain breakage results in the removal of radioisotopes used during $\gamma$ radioassay, then assay results will lead to an underrepresentation of the total number of radionuclides present in the material.

In this work, we quantify the effects of primordial radioisotope chain disequilibrium on neutron and $\gamma$ emission. Monte Carlo is used to estimate the impact of these changes on a liquid Xe detector similar to the current LUX and XENON TPC designs \cite{luxnim,xe100,lz,xe1t}. A model liquid Ar detector of equal mass, similar to the ArDM and DarkSide designs \cite{ardm,darkside}, is used to estimate the impact of chain disequilibrium on Ar targets.

\section{\label{Screening-methods}Screening Methodologies}

The most common method for measuring radioisotope concentrations in construction materials is to expose material samples to a high-sensitivity Ge detector \cite{luxnim,xe100ms,next,kurf,gerda}. This method yields a direct measurement of the $\gamma$ energy spectrum emitted by all radioisotopes in the material. This method is well-suited to screening large quantities of construction materials in a single screening run, which is crucial for experiments which use very large quantities of these materials in close proximity to their active targets. This also enables $\gamma$ screeners to average over any variations in radioisotope concentration with position in the sample. Gamma screening yields comparable sensitivity per $\gamma$ emitted for all radioisotopes. For these reasons, $\gamma$ radioassay is widely used as the primary radioassay method for material screening programs in low background experiments.

Isotopes with low $\gamma$ emission probability per decay are identified with relatively high upper limits. This means that $\gamma$ screening is relatively insensitive to several radioisotopes shown in the decay chains in Fig.~\ref{238U-decay-chain} and \ref{235U-232Th-decay-chains}. Typical $\gamma$ detectors also have backgrounds which fall exponentially with increasing energy, with an order of magnitude or greater change in background rate between 100~keV and 1000~keV. This generally leads to the preferential use of high-energy $\gamma$ lines for radioisotope identification. Standard identification of the $^{238}$U and $^{232}$Th chains uses $\gamma$ lines $>$300~keV.

Mass spectrometry techniques can also be used to identify radioisotopes in material samples \cite{xe100ms,next}. In principle, this technique can be useful for distinguishing radioisotopes which have a low $\gamma$ emission intensity, and would therefore escape detection by Ge screeners. However, these techniques are limited in their ability to detect daughter nuclei in long decay chains. In secular equilibrium, the ratios of daughter isotope to head isotope half-lives determines the concentrations of the daughter isotopes. In the case of the decay chains in Fig. \ref{238U-decay-chain} and \ref{235U-232Th-decay-chains}, the parent half-lives are higher than those of the daughters by 4--25 orders of magnitude. Mass spectrometry can therefore set meaningful upper limits only on the head isotopes of the decay chains. Mass spectrometry also samples very small quantities of material in a given run, which leads to potential error in inhomogeneous samples.

\section{\label{Chain-Disequilibrium-Definition}Chain Disequilibrium for Primordial Radionuclides}

\subsection{\label{238U}$^{238}$U Chain}

In principle, disequilibrium can be introduced at any stage in a decay chain. However, disequilibrium is only relevant to dark matter experiments in the case that disequilibrium occurs by removing an isotope with a half-life comparable to or greater than the timescale of the experiment. This timescale is typically measured in years. From Fig.~\ref{238U-decay-chain} and \ref{235U-232Th-decay-chains}, the decay chain most likely to yield a relevant break in equilibrium is that of $^{238}$U, which has several isotopes with half-lives significantly longer than this limit. For this work, we define three $^{238}$U sub-chains based on the relatively long half-lives of their parent isotopes: the $^{238}$U~early sub-chain, including isotopes from $^{238}$U to $^{230}$Th; the $^{226}$Ra sub-chain, including isotopes from $^{226}$Ra to $^{214}$Po; and the $^{210}$Pb sub-chain, including $^{210}$Pb and its daughters. 

The definition of the $^{226}$Ra sub-chain is conservative for this study. If breakage occurs below $^{226}$Ra, the relatively short half-lives of the daughters (set by $^{222}$Rn at 3.8~days) will lead to re-establishment of equilibrium well before radioassay measurements are performed. Breakage above $^{226}$Ra will lessen the effect of disequilibrium on detector backgrounds, since there will be fewer unaccounted radioisotopes in excess. Setting $^{226}$Ra as the parent of the middle sub-chain therefore leads to the greatest potential for impact of the sub-chain on detector backgrounds. The specific details of equilibrium breakage are dependent on the chemical processes used in the treatment of materials. We do not address these processes in this work, but instead make simple conservative assumptions about the disequilibrium scenario.

Although $\gamma$~rays from the $^{238}$U~early sub-chain have considerably lower branching ratios than those from the $^{226}$Ra sub-chain, typical $\gamma$ counters can still place useful upper limits on $^{238}$U~early sub-chain activity. The ratio of $^{238}$U~early sub-chain limits to $^{226}$Ra sub-chain limits set by $\gamma$ counting is comparable to the ratio of the intensities of the $\gamma$~rays used to identify these sub-chains. For the $^{226}$Ra sub-chain, the typical energies used for identification are the 352~keV and 609~keV lines, with branching ratios of 36\% and 47\% respectively. The $^{238}$U~early sub-chain is generally identified by the 1.0~MeV line from $^{234}$Pa, which has an intensity a factor $\times$1/60 and $\times$1/80 that of the 352~keV and 609~keV lines, respectively. Variability in detection efficiency and detector backgrounds can result in a wide range of upper limits for the $^{238}$U~early sub-chain relative to the $^{226}$Ra sub-chain, with typical ratios ranging from $\times$20--60 \cite{xe100ms,next,kurf,gerda,r11410,luxti}.

The $^{238}$U chain contains one long-lived radioisotope after the $^{226}$Ra sub-chain, $^{210}$Pb, with a half-life of 22~years. The $^{210}$Pb chain contains no high-energy $\gamma$~rays with high branching ratios, so the expected increase in detector ER backgrounds is minimal. However, one of the $^{210}$Pb sub-chain daughters undergoes $\alpha$ decay, potentially leading to neutron generation. This work also considers neutron generation from the $^{210}$Pb sub-chain.

It should be noted that $^{210}$Pb and its daughters do not produce $\gamma$~rays which are easily observable in typical radioassay Ge detectors. Gamma emission comes primarily from $^{210}$Pb decay, with a 4\% branching ratio for generation of a 46~keV $\gamma$. The mean free path of this $\gamma$ is 4~cm in H$_2$O. The $\gamma$ is easily absorbed within bulk samples before reaching the Ge detector, thereby heavily suppressing its signal and greatly raising upper limits on its activity. Gamma screening backgrounds are also typically high in the energy range around 46~keV, further weakening upper limit activity measurements. Screening results for $^{210}$Pb are therefore typically not reported in $\gamma$ radioassays. We do not attempt to quantify likely concentrations of the $^{210}$Pb sub-chain in this work.

\subsection{\label{235U}$^{235}$U Chain}

The concentration of $^{235}$U in nature is 0.7\% that of $^{238}$U. Standard chemical processes do not alter the $^{235}$U / $^{238}$U ratio in materials. It is therefore common to assume that screened materials have a natural ratio of $^{235}$U / $^{238}$U, which sets a much lower limit on the amount of $^{235}$U than does direct counting. Since $^{235}$U and its daughters are not directly measured, $^{235}$U chain disequilibrium is not addressed in this work.

If $\gamma$ screening misidentifies the amount of $^{238}$U in a material due to a factor $\times$1/60 deficit in the $^{226}$Ra sub-chain, then the amount of $^{235}$U will be misidentified by the same factor. In this scenario, the $^{235}$U decay rate could potentially be 50\% of the $^{226}$Ra sub-chain decay rate. However, the $^{235}$U chain has one less $\alpha$ emission per parent decay than the $^{238}$U chain, a comparable mean $\alpha$ energy, and lower-energy $\gamma$~rays with lower intensities. $^{235}$U will contribute neutron and $\gamma$ backgrounds no larger than those already present from existing $^{226}$Ra in the experiment, as discussed in Sec. \ref{Impact-on-DM-Neutrons} and \ref{Impact-on-DM-Gammas}. It is also notable that, in this scenario, the $^{235}$U chain would be detectable in $\gamma$ radioassay by the 185~keV $\gamma$ line from $^{235}$U decay.

\subsection{\label{232Th}$^{232}$Th Chain}

The $^{232}$Th chain can only be broken on a timescale of years, matching the common lifetime of current and future dark~matter experiments. The longest-lived daughter below $^{232}$Th is $^{228}$Ra, with a half-life of 5.8~years. If $^{228}$Ra and its daughters are completely removed from a material, then the rate of $^{228}$Ra decays will reach a factor $\times$1/2 that of $^{232}$Th after 6~years. $^{228}$Th follows two generations after $^{228}$Ra, with a half-life of 1.9~years. In the scenario where all $^{232}$Th daughters have been removed, $^{228}$Th reaches a decay rate a factor $\times$1/2 that of $^{232}$Th after 9~years. The longest-lived daughter after $^{228}$Th has a half-life on the order of minutes; the decay rates of all of the daughters thus closely shadow that of $^{228}$Th.

Chain disequilibrium could also be introduced by the removal of only $^{228}$Th and its daughters. In this case, chain equilibrium would be re-established within a factor $\times$1/2 after 2~years. This would also leave the isotope $^{228}$Ac, which generates several high-energy $\gamma$ rays easily detectable by $\gamma$ radioassay. The most conservative assumption for $^{232}$Th chain disequilibrium is therefore the removal of $^{228}$Ra and its daughters, as this introduces the most appreciable lag in the re-establishment of chain equilibrium with a low chance of detection.

If a material were to be processed with a method that removed all $^{232}$Th daughters and left the original $^{232}$Th, then the timescale on which $^{232}$Th / $^{228}$Ra / $^{228}$Th equilibrium is re-established would lead to a background source which increases significantly over the experiment lifetime. Consider a material which undergoes complete removal of all $^{232}$Th daughters, and is then subject to radioassay 45~days afterward. After 45~days, the rate of $^{228}$Ra decays (closely traced by detectable $^{228}$Ac) would be only 1\% the rate of $^{232}$Th decays. The reported $^{232}$Th rate from $\gamma$ radioassay, which uses lines from $^{228}$Ac and $^{228}$Th daughters, would then be $\times$1/100 the true $^{232}$Th rate. The $^{228}$Ra and $^{228}$Th rates grow linearly with time, increasing over an order of magnitude over the experiment lifetime.

Fortunately, removal of all $^{232}$Th daughters by chemical processing is unrealistic. $^{232}$Th is chemically identical to $^{228}$Th; any process which removes $^{228}$Th would remove $^{232}$Th in comparable quantities. If processing left $^{228}$Th in the material, then $^{228}$Th and its daughters would reach equilibrium within days. Gamma radioassay would then detect the $^{228}$Th daughter signatures, and the inferred $^{232}$Th quantity would be correct. Similar arguments apply to a scenario in which Th is introduced into a material during processing. It is notable that detection of a deficit of $^{228}$Ac relative to $^{228}$Th daughters would indicate that processing techniques had induced disequilibrium in the isotope chain before counting. This could serve as an indicator for potential disequilibrium conditions in the $^{238}$U chain in the same material, as processes which removed Ra isotopes and left Th isotopes would be identified.

\section{\label{Alpha-neutron-yields}$^{238}$U $\alpha$ Emission Spectra and Neutron Yields}

Experiments searching for low-energy NR signals are particularly sensitive to backgrounds from neutron scattering. With sufficient external shielding material, neutron generation from radionuclides in detector materials contributes the dominant NR rate in the target volume. Neutron generation from detector materials comes primarily from spontaneous fission ($^{238}$U having the most significant contribution through this channel), and from ($\alpha$,n) interactions. Disequilibrium in primordial radionuclide chains can lead to an underestimation of $\alpha$-generating isotopes, leading in turn to an underestimation of neutron yields from detector materials.

\begin{figure}
\begin{centering}
\includegraphics[width=1\textwidth]{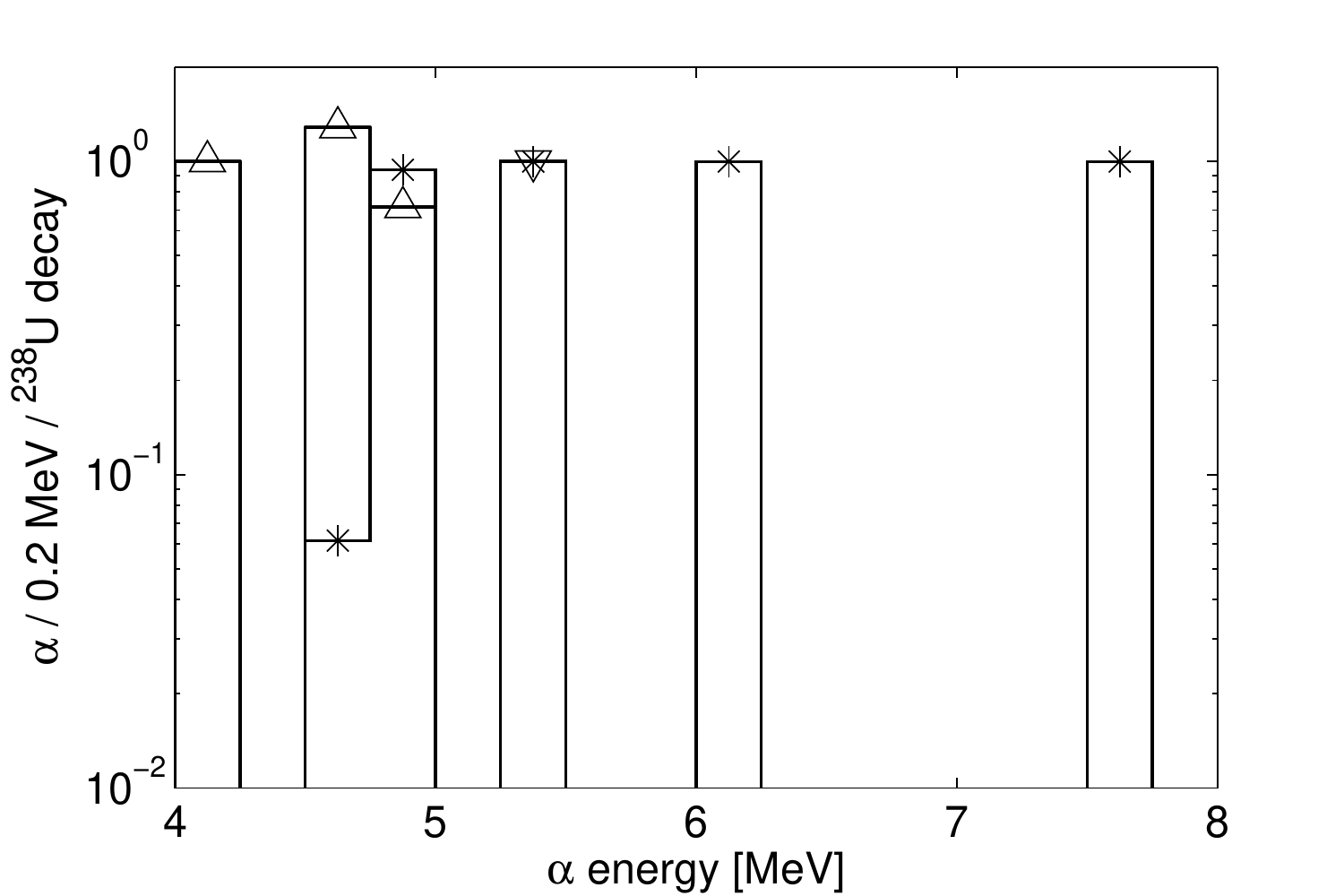}
\par\end{centering}
\caption{\label{Alpha-energies}Alpha energies generated from $^{238}$U decay chain isotopes, assuming full secular equilibrium. Alphas are shown corresponding to the $^{238}$U~early sub-chain (upward arrows), $^{226}$Ra sub-chain (stars), and $^{210}$Pb sub-chain (downward arrows). Data is taken from \cite{nndc}.}
\end{figure}

Neutron yields vary by orders of magnitude across the spectrum of $\alpha$ energies associated with $^{238}$U chain isotopes, as described in a review of ($\alpha$,n) yields by Heaton et al. \cite{heaton}. The neutron rate from a given sub-chain convolves neutron yield as a function of $\alpha$ energy with the $\alpha$ energies of each sub-chain. For $^{238}$U isotopes, $\alpha$ energy spectra are shown in Fig.~\ref{Alpha-energies}.

\begin{figure}
\begin{centering}
\includegraphics[width=1\textwidth]{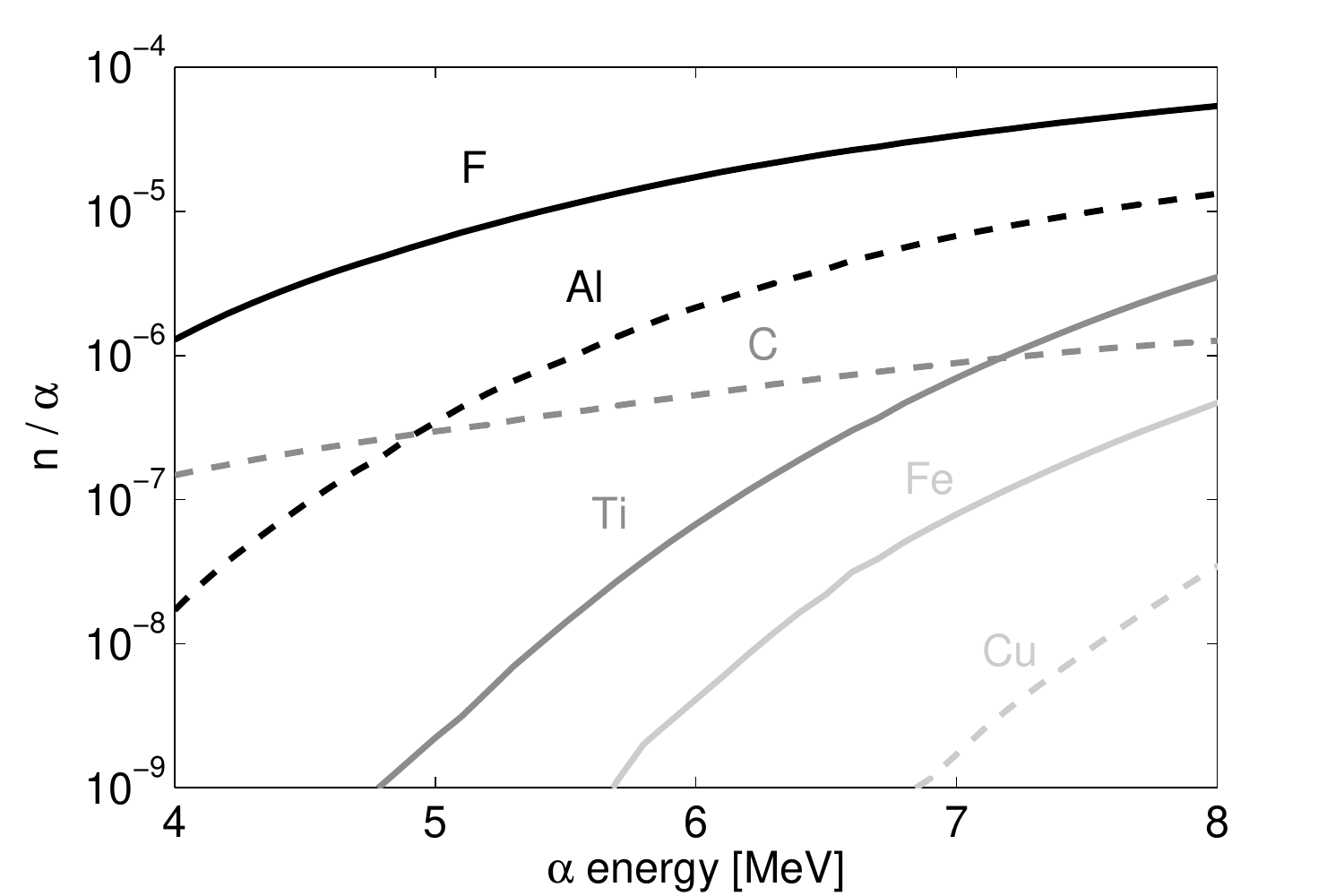}
\par\end{centering}
\caption{\label{n_per_a_calc}Neutrons produced per incident $\alpha$ particle for various common elements in detector construction materials, as a function of $\alpha$ kinetic energy. Lines are shown for F (dark solid), Al (dark dashed), Ti (medium solid), C (medium dashed), Fe (light solid), and Cu (light dashed).}
\end{figure}

The neutron spectrum as a function of initial $\alpha$ energy $E_{\alpha,0}$ for material comprised of an element with molar mass $A$ is calculated as
\begin{equation}
Y\left(E_{\alpha,0},E_{n}\right)=\frac{N_{A}}{A}\int_{0}^{E_{\alpha,0}}\frac{\sigma\left(E_{\alpha},E_{n}\right)}{S_\alpha\left(E_{\alpha}\right)}dE_{\alpha},\label{n-yield-per-a}
\end{equation}
where $Y\left(E_{\alpha,0},E_{n}\right)$ is the neutron yield per incident $\alpha$ particle with energy $E_{\alpha,0}$; $\sigma\left(E_{\alpha},E_{n}\right)$ is the ($\alpha$,n) cross section for incoming $\alpha$ energy $E_\alpha$ and outgoing neutron energy $E_n$; and $S_\alpha\left(E_{\alpha}\right)$ is the $\alpha$ stopping power for the material. The methodology follows that described in \cite{mei}. The ($\alpha$,n) cross sections were calculated for several elements using the TALYS 1.4 simulation program \cite{talys}. $S_\alpha$ data were taken from the ASTAR database \cite{astar}. Integrated neutron production rates were compared to those listed in Heaton and found to match within a factor of two, with an average overproduction by a factor of 20\%. The exception is C, for which the neutron yield is calculated to be an average $\times$5 higher than that reported in Heaton. The TALYS cross section results for C match measurements reported in \cite{harissopulos}, and are used for this work. Yield curves integrated over all neutron energies are shown in Fig.~\ref{n_per_a_calc}.

\begin{figure}
\begin{centering}
\includegraphics[width=1\textwidth]{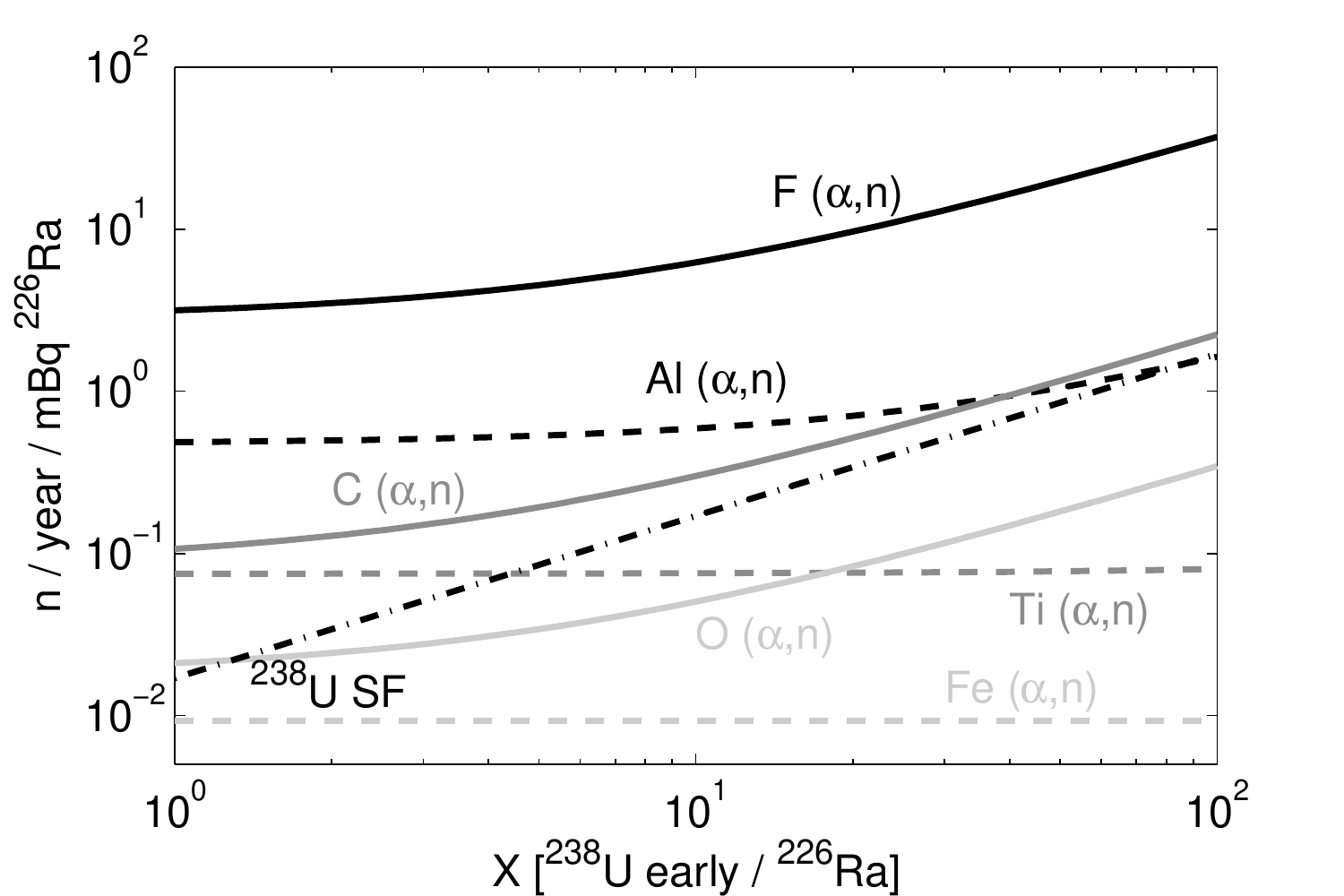}
\par\end{centering}
\caption{\label{n-rate-vs-chain-mult}Neutron rate from ($\alpha$,n) processes per $^{226}$Ra sub-chain decay rate, as a function of $^{238}$U~early sub-chain / $^{226}$Ra sub-chain ratio $X$, for various materials. Overlaid is the yield from $^{238}$U spontaneous fission, assuming that fission yields single neutrons which cannot be vetoed by accompanying neutrons or $\gamma$ rays. Lines are shown for F (dark solid), Al (dark dashed), Ti (medium solid), C (medium dashed), Fe (light solid), Cu (light dashed), and spontaneous fission (dark dash-dotted). Typical $\gamma$ radioassay measurements limit the potential $^{238}$U~early fraction excess to $\times$20--60. }
\end{figure}

We define the ratio of $^{238}$U~early sub-chain to $^{226}$Ra sub-chain $X$. We then define the ($\alpha$,n) neutron yield multiplier from chain equilibrium $\Gamma\left(X,E_{n}\right)$, which is the fractional increase in yield for neutrons with energy $E_n$ over the case of full-chain equilibrium, as a function of $X$. $X$=1 corresponds to $^{238}$U chain equilibrium. $\Gamma\left(X,E_{n}\right)$ is calculated as
\begin{equation}
\Gamma\left(X,E_{n}\right)=\frac{\left(X-1\right)\left(\sum_{j}Y\left(E_{\alpha,0,j},E_{n}\right)\right)}{\sum_{k}Y\left(E_{\alpha,0,k},E_{n}\right)}+1,\label{n-yield-vs-chain-excess}
\end{equation}
where $j$ and $k$ are iterators spanning over the $\alpha$ particles generated from the $^{238}$U~early sub-chain and full $^{238}$U decay chain in equilibrium, respectively. Fig.~\ref{n-rate-vs-chain-mult} shows the total neutron generation rate integrated over all neutron energies, $\int\Gamma\left(X,E_{n}\right)dE_{n}$, as a function of $X$ for several common elements in detector construction materials. The summed neutron yield per sub-chain decay for all $^{238}$U sub-chains is given in Table \ref{subchain-a-n-yields}.

Shown separately in Fig.~\ref{n-rate-vs-chain-mult} is the $^{238}$U spontaneous fission rate, equal to $5.4\times10^{-7}$~fissions per $^{238}$U decay \cite{toi}. For materials with a low neutron yield below 5~MeV $\alpha$ energy, fission gives the dominant neutron yield in the case of $^{238}$U~early sub-chain excess. It should be noted that an extremely conservative model is used where each spontaneous fission event contributes a single neutron, with no other prompt emission. In reality, spontaneous fission promptly releases multiple neutrons and/or $\gamma$ rays. The chance of veto of these events is extremely high due to multiple-scatter and high-energy event rejection in many dark matter detectors. 

\begin{figure}
\begin{centering}
\includegraphics[width=1\textwidth]{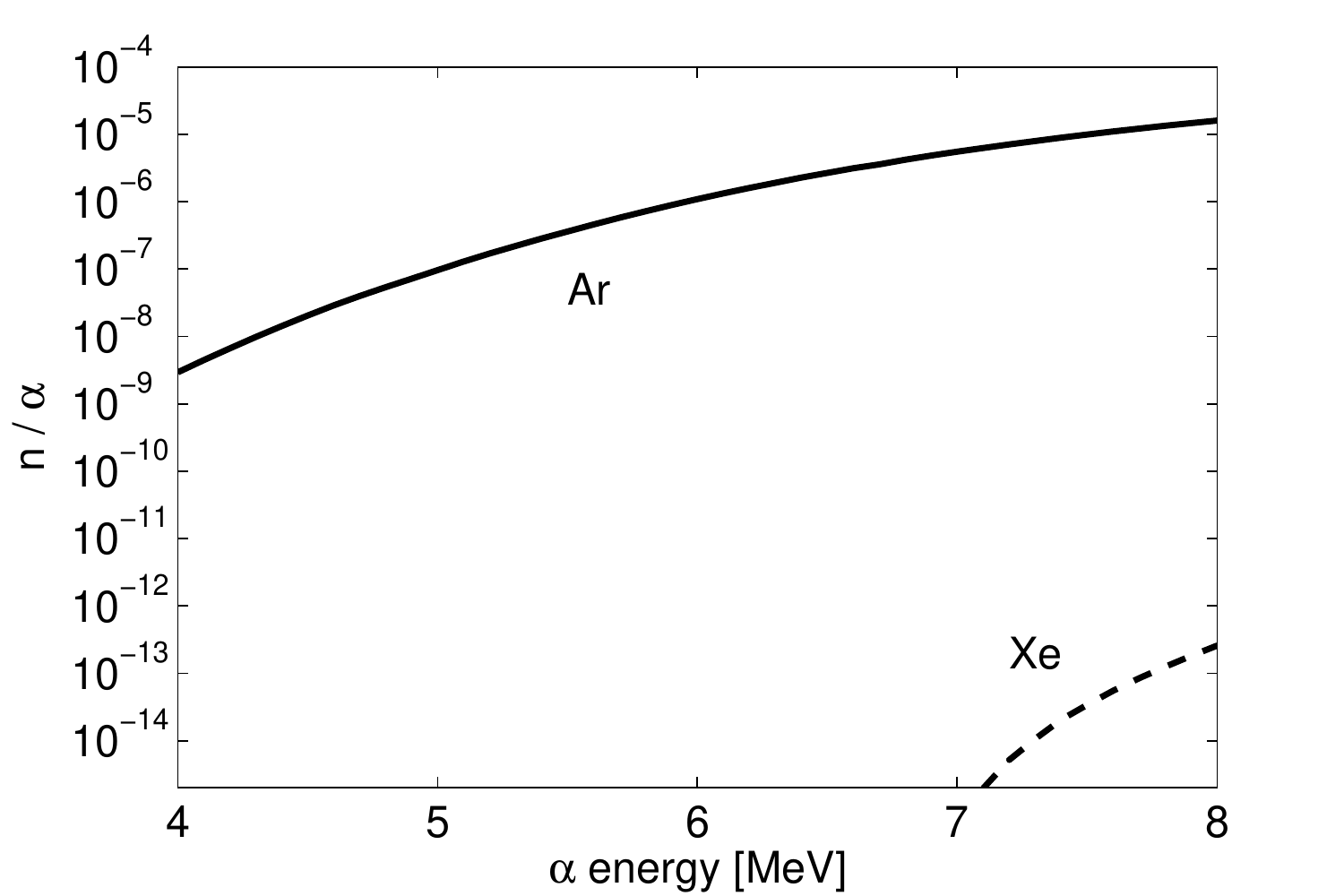}
\par\end{centering}
\caption{\label{n_per_a_target}Neutrons produced per incident $\alpha$ particle for Ar (solid) and Xe (dashed), as a function of $\alpha$ kinetic energy.}
\end{figure}

It is interesting to note the ($\alpha$,n) yields for direct incidence of $\alpha$ particles on Ar and Xe target materials. The yield curves as a function of incident $\alpha$ energy are shown in Fig.~\ref{n_per_a_target}. The Ar target is sensitive to all $\alpha$ energies found in the $^{238}$U and $^{232}$Th chains. A particular risk is that of neutron production from $^{222}$Rn leakage into the Ar target material, which can occur from $^{226}$Ra sources on the surface of detector materials. The presence of $^{222}$Rn and its daughters results in a neutron generation rate of 470~n~yr$^{-1}$~(Bq~$^{222}$Rn)$^{-1}$. Xe targets are vulnerable only to the 7.8~MeV $\alpha$ from the $^{222}$Rn decay chain, with a yield of $3.8\times10^{-6}$~n~yr$^{-1}$~(Bq~$^{222}$Rn)$^{-1}$. Typical self-shielding cuts would have reduced effectiveness against this background due to the internal generation of the neutrons and the consequent reduction in path length required for the neutrons to escape.

Surface deposition of $^{222}$Rn progeny on detector surfaces can also result in a significant $^{210}$Pb concentration exposed directly to the target material. $^{210}$Pb $\alpha$ particles yield 3.5~n~yr$^{-1}$~(Bq~$^{210}$Pb)$^{-1}$ on Ar, assuming 50\% of $\alpha$ particles are emitted into detector surfaces and do not contribute to Ar ($\alpha$,n). $^{210}$Pb $\alpha$ interactions in Xe do not generate neutrons, as Xe ($\alpha$,n) processes have an $\alpha$ energy threshold well above 5.3~MeV.

\begin{table}
\begin{centering}
\begin{tabular}{r|ccc}
\multicolumn{1}{c|}{} 	& $^{238}$U~early 		& $^{226}$Ra 	& $^{210}$Pb \\
\hline
Al 				& 37 				& 1400 	& 66 \\
C 				& 68 				& 240 	& 35 \\
Cu 				& -- 				& 16 		& -- \\
F 				& 1100 			& 8000 	& 890 \\
Fe 				& 0.00030		 	& 30 		& 0.0015 \\
O 				& 10 				& 49 		& 6.9 \\
Ti 				& 0.17 			& 240 	& 0.69 \\
\end{tabular}
\par\end{centering}
\caption{\label{subchain-a-n-yields}$^{238}$U sub-chain ($\alpha$,n) neutron yields per parent decay, multiplied by 10$^8$, for common detector construction materials. Yields are calculated from $\alpha$ energies in Fig.~\ref{Alpha-energies} and ($\alpha$,n) neutron yields in Fig.~\ref{n_per_a_calc}. Yields from the $^{238}$U~early sub-chain do not include contributions from $^{238}$U spontaneous fission, the probability for which is $5.4\times10^{-7}$.}
\end{table}

\section{\label{Impact-on-DM-Neutrons}$^{238}$U Disequilibrium Impact on Neutron Backgrounds for Dark Matter Experiments}

\begin{figure}
\begin{centering}
\includegraphics[width=1\textwidth]{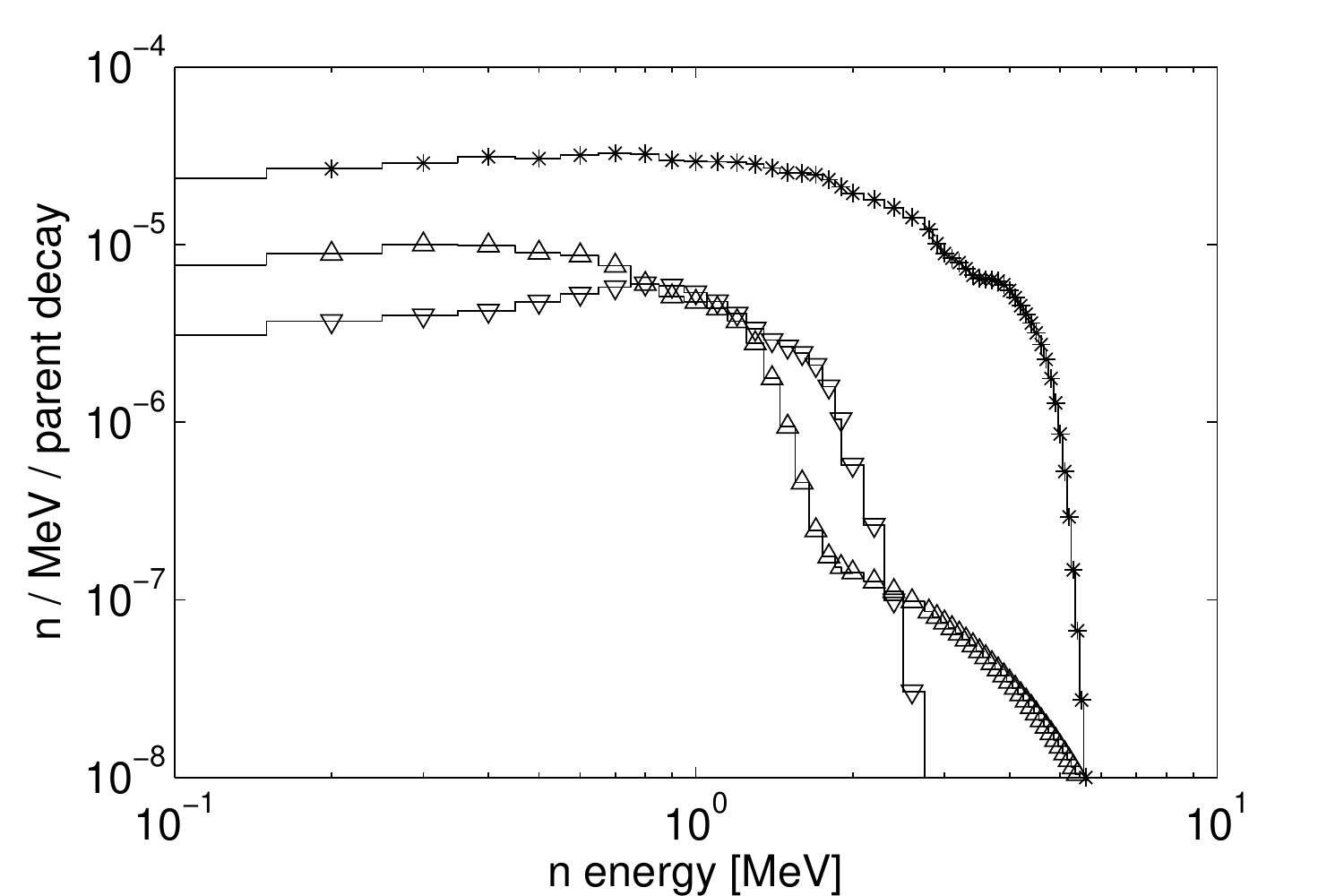}
\par\end{centering}
\caption{\label{n_E_spec}Neutron emission energy spectra corresponding to $^{238}$U spontaneous fission and ($\alpha$,n) emission on F from the $^{238}$U~early sub-chain (upward arrows), $^{226}$Ra sub-chain (stars), and $^{210}$Pb sub-chain (downward arrows). Spectra were calculated from Eq. \ref{n-yield-per-a}.}
\end{figure}

The impact of neutrons on dark~matter sensitivity was assessed by direct comparison of projected NR rates from neutrons and WIMPs. The example of fluoridated material (e.g. PTFE, or Teflon\textregistered) was chosen for this study due to its relatively high neutron yield over other common detector construction materials. The neutron energy spectrum from ($\alpha$,n) on F was calculated from Eq. \ref{n-yield-per-a} for the $^{238}$U~early sub-chain, $^{226}$Ra sub-chain, and $^{210}$Pb sub-chain. The $^{238}$U spontaneous fission neutron spectrum was added to the $^{238}$U~early chain. The spectra are shown in Fig.~\ref{n_E_spec}.

\begin{figure}
\begin{centering}
\subfloat[]{\includegraphics[width=0.5\textwidth]{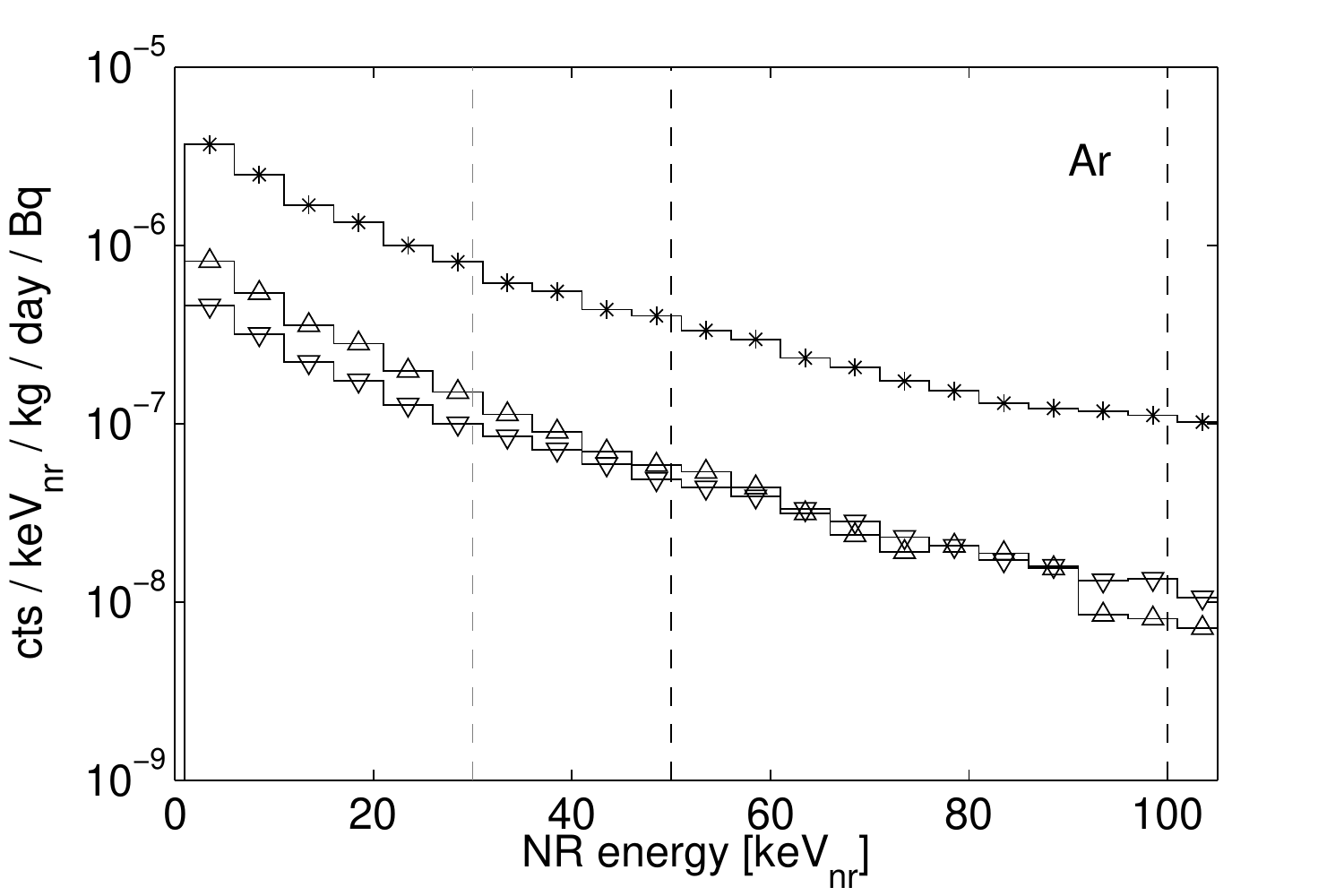}
}
\subfloat[]{\includegraphics[width=0.5\textwidth]{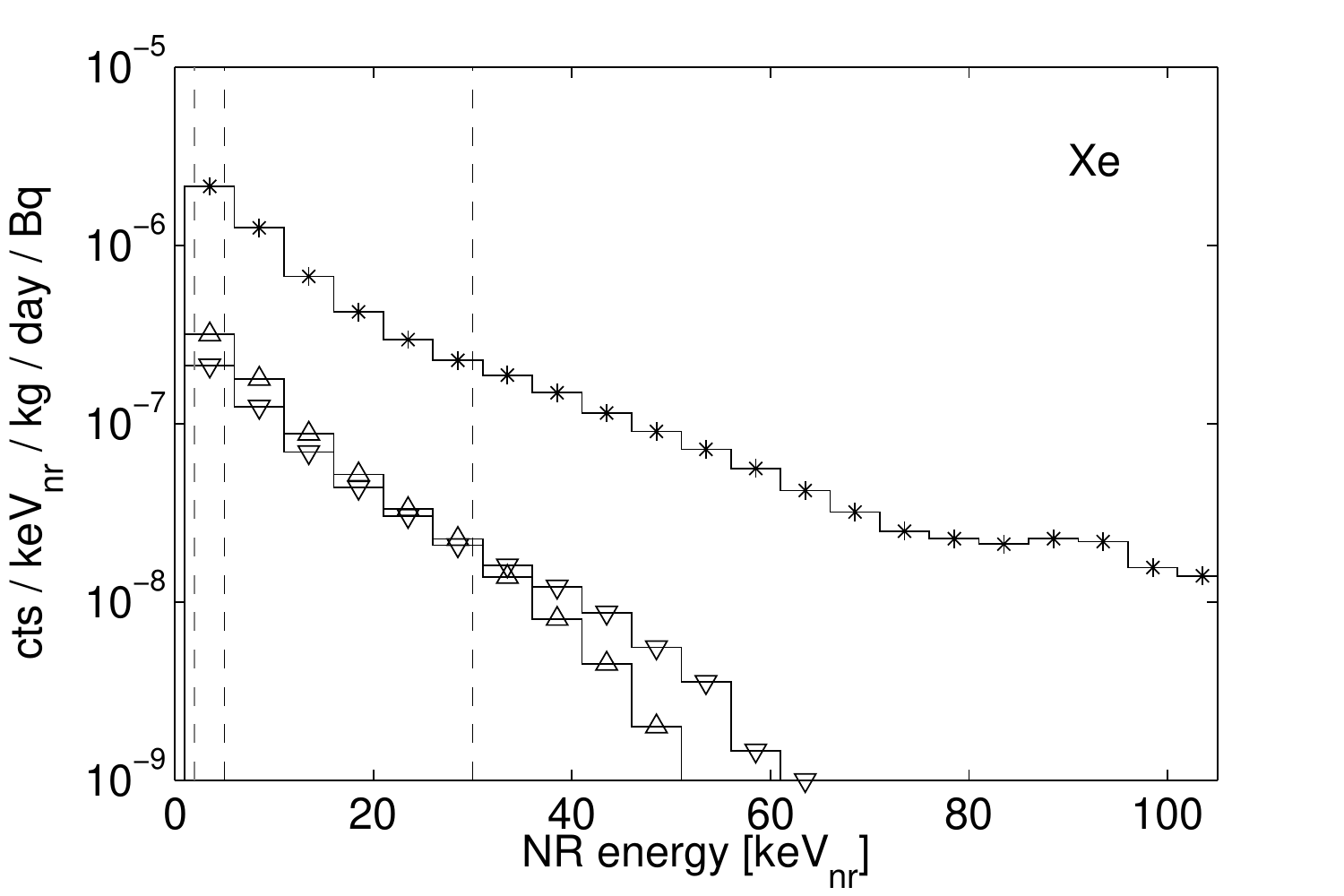}
}
\par\end{centering}
\caption{\label{NR_spec}NR spectra in simulated one~tonne (a) liquid Ar and (b) liquid Xe detectors, using geometry as described in Sec. \ref{Impact-on-DM-Gammas}. Recoil spectra correspond to neutrons generated from the $^{238}$U~early sub-chain (upward arrows), $^{226}$Ra sub-chain (stars), and $^{210}$Pb sub-chain (downward arrows), with incident neutron energy spectra given in Fig.~\ref{n_E_spec}. Black dashed vertical bars indicate typical WIMP search windows for both detectors, with gray bars indicating potential improvements in low-energy threshold. Rates are found in a 500~kg cylindrical fiducial volume. A cut removing multiple scatter events is applied.}
\end{figure}

The impact of $^{238}$U chain disequilibrium is assessed by Monte Carlo for one~tonne ideal liquid Ar and liquid Xe dark~matter detectors. The simulations are based on the Geant4 toolkit \cite{geant}. The detectors have a cylindrical geometry with 1:1 aspect ratio, and are surrounded by a cylindrical water shield of thickness 1~m. Neutrons are thrown from the border between the target material and the water shield, representing activity from detector internals. The neutron scatter rate is then found in a cylindrical 500~kg fiducial region with 1:1 aspect ratio. A single-scatter cut is applied, requiring events to have an energy-weighted standard deviation of $<$2~cm in radius and $<$0.5~cm in height, comparable to the typical spatial resolution for Xe TPC detectors \cite{xe100,luxsr}. Single-scatter cut efficiency is insensitive to the exact length threshold used. Rates are plotted as a function of recoil energy in Fig.~\ref{NR_spec}, normalized per neutron emitted.

\begin{figure}
\begin{centering}
\subfloat[]{\includegraphics[width=0.5\textwidth]{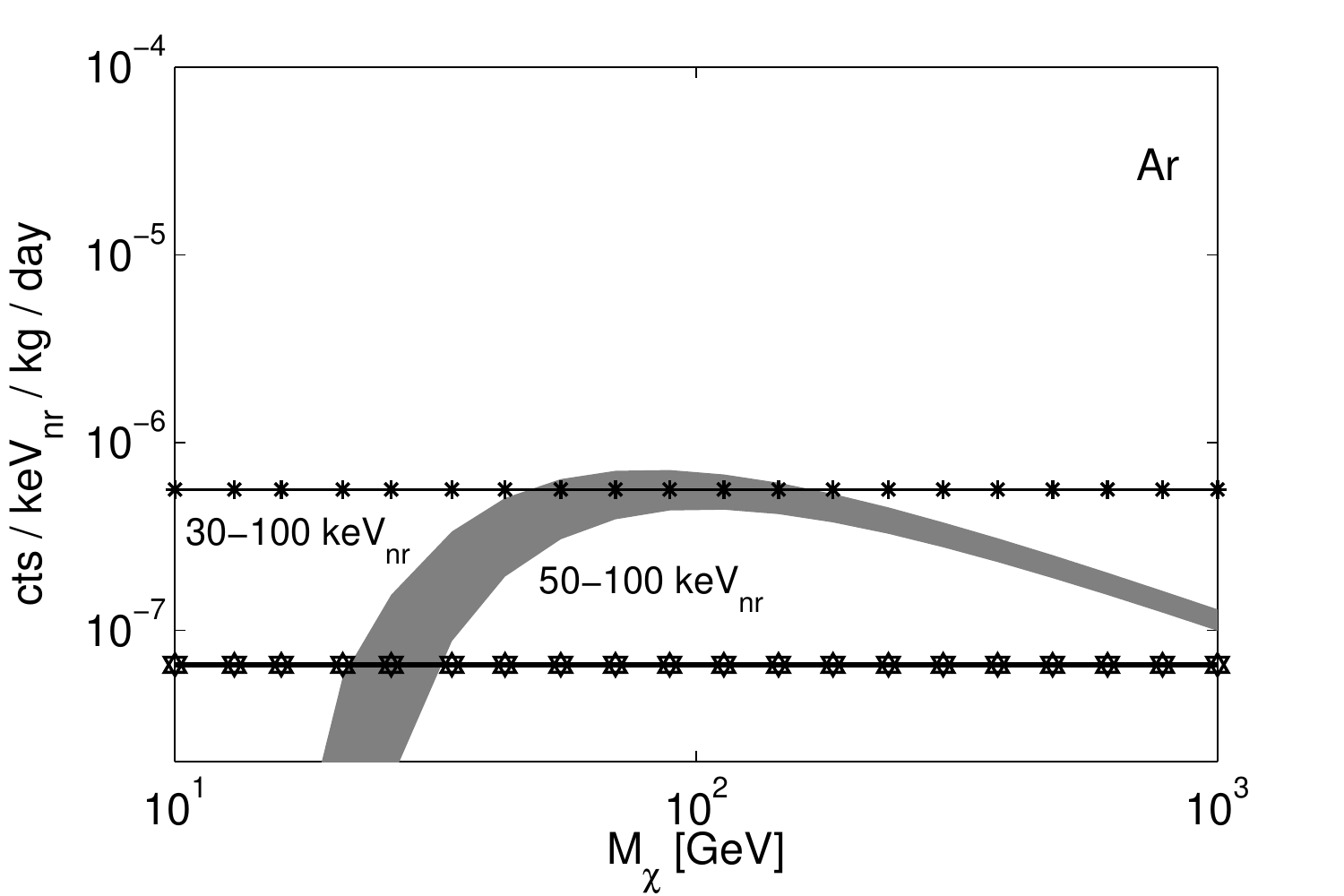}
}
\subfloat[]{\includegraphics[width=0.5\textwidth]{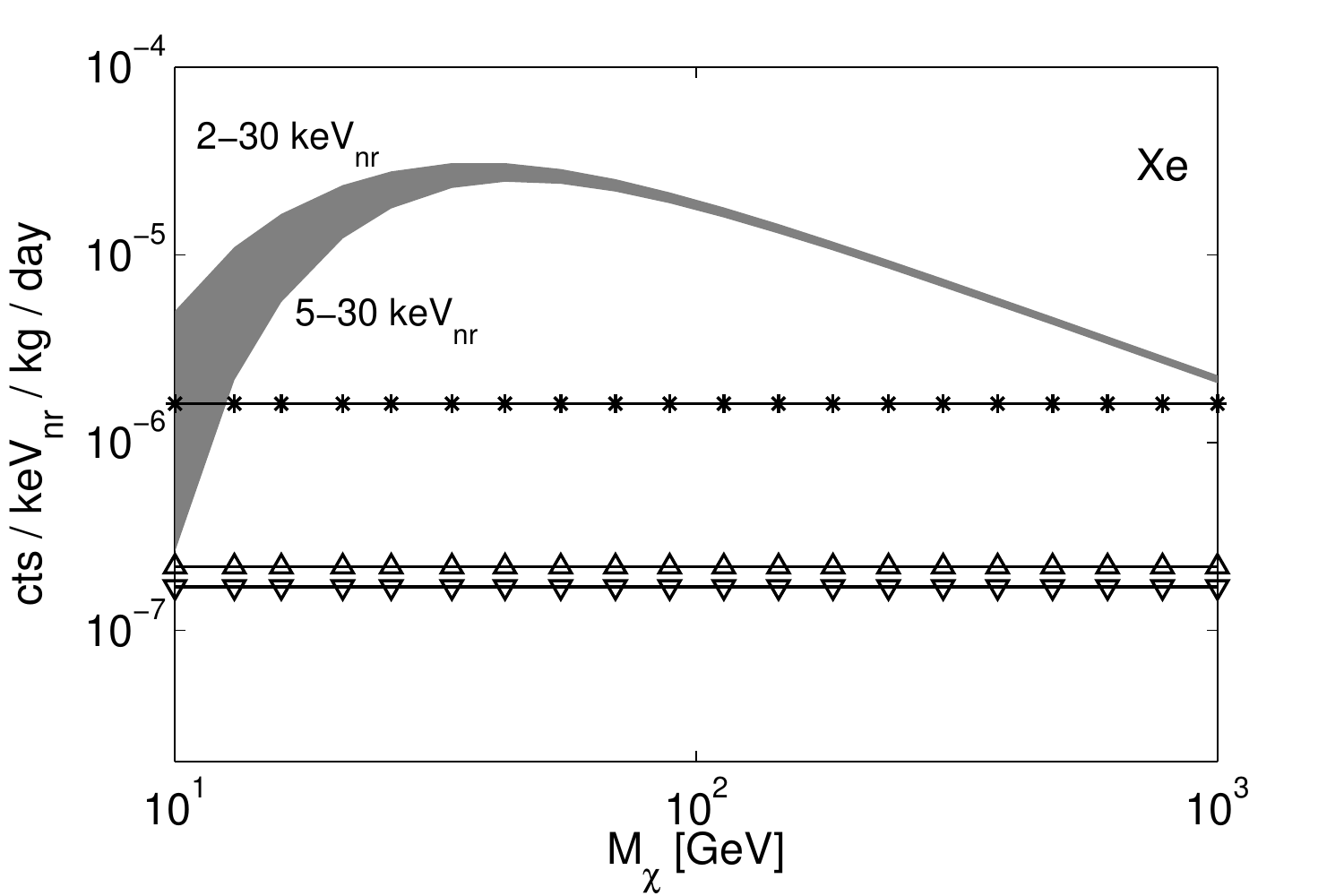}
}
\par\end{centering}
\caption{\label{wimp_rate_vs_mass}WIMP signal and neutron rates as a function of WIMP mass $M_\chi$ in simulated one-tonne (a) liquid Ar and (b) liquid Xe detectors. A WIMP-nucleon spin-independent cross-section of $10^{-45}$~cm$^2$ is assumed. Neutron rates correspond to 1~Bq of $^{238}$U~early sub-chain (upward arrows), $^{226}$Ra sub-chain (stars), and $^{210}$Pb sub-chain (downward arrows). Note that $^{238}$U~early sub-chain and $^{210}$Pb sub-chain NR rates overlap for the Ar detector, at $7\times10^{-8}$. Rates are calculated using NR energy windows of 50-100~keV$_{nr}$ for Ar and 5-30~keV$_{nr}$ for Xe. WIMP rates are shown as bands illustrating the effect of lowering the detection threshold to 30~keV$_{nr}$ for Ar and 2~keV$_{nr}$ for Xe, as shown in Fig.~\ref{NR_spec}.}
\end{figure}

The projected WIMP signal as a function of WIMP mass $M_\chi$ is shown in Fig.~\ref{wimp_rate_vs_mass} for both Ar and Xe detectors. A WIMP-nucleon spin-independent interaction cross-section of $10^{-45}$~cm$^2$ is used. Overlaid are neutron scatter rates for each of the $^{238}$U sub-chains, after application of analysis cuts. A parent rate of 1~Bq is assumed for each sub-chain. WIMP and neutron rates are calculated from Fig.~\ref{NR_spec} in the energy range 50-100~keV$_{nr}$ for the Ar detector \cite{miniclean}, and 5-30~keV$_{nr}$ for the Xe detector \cite{luxnim,xe100,xe10}, where keV$_{nr}$ is defined as energy deposited in the detector by NR. WIMP rates are shown as bands which illustrate the projected change in sensitivity if the lower bound of the WIMP search window is lowered to 30~keV$_{nr}$ for the Ar detector \cite{ardm} and 2~keV$_{nr}$ for the Xe detector \cite{xe10lm}. Neutron rates increase by $<$$\times$2 in these scenarios.

For both Ar and Xe detectors, the $^{238}$U~early sub-chain contributes $\times$1/10 the neutron rate of the $^{226}$Ra sub-chain. The increase closely tracks the raw neutron emission rate as a function of $^{238}$U~early chain imbalance in Fig.~\ref{n-rate-vs-chain-mult}, indicating insensitivity to the exact shape of the neutron energy spectra in Fig.~\ref{n_E_spec}. The neutron rates in the two detectors are comparable in their respective WIMP search energy regions. The Xe detector takes advantage of its low energy threshold for a high WIMP detection rate. The WIMP signal rate in the Xe detector is a factor $\times$12 above background and a factor $\times$40 above the WIMP rate in the Ar detector, at $M_\chi=100$~GeV. The Xe detector shows greater resilience against neutron backgrounds due to its higher WIMP signal rate, and is able to achieve comparable discovery potentials to the Ar detector with an order of magnitude higher background, or detect WIMPs with an order of magnitude lower interaction cross-section at comparable background levels.

\section{\label{Impact-on-DM-Gammas}$^{238}$U Disequilibrium Impact on $\gamma$ Backgrounds for Dark Matter Experiments}

\begin{figure}
\begin{centering}
\includegraphics[width=1\textwidth]{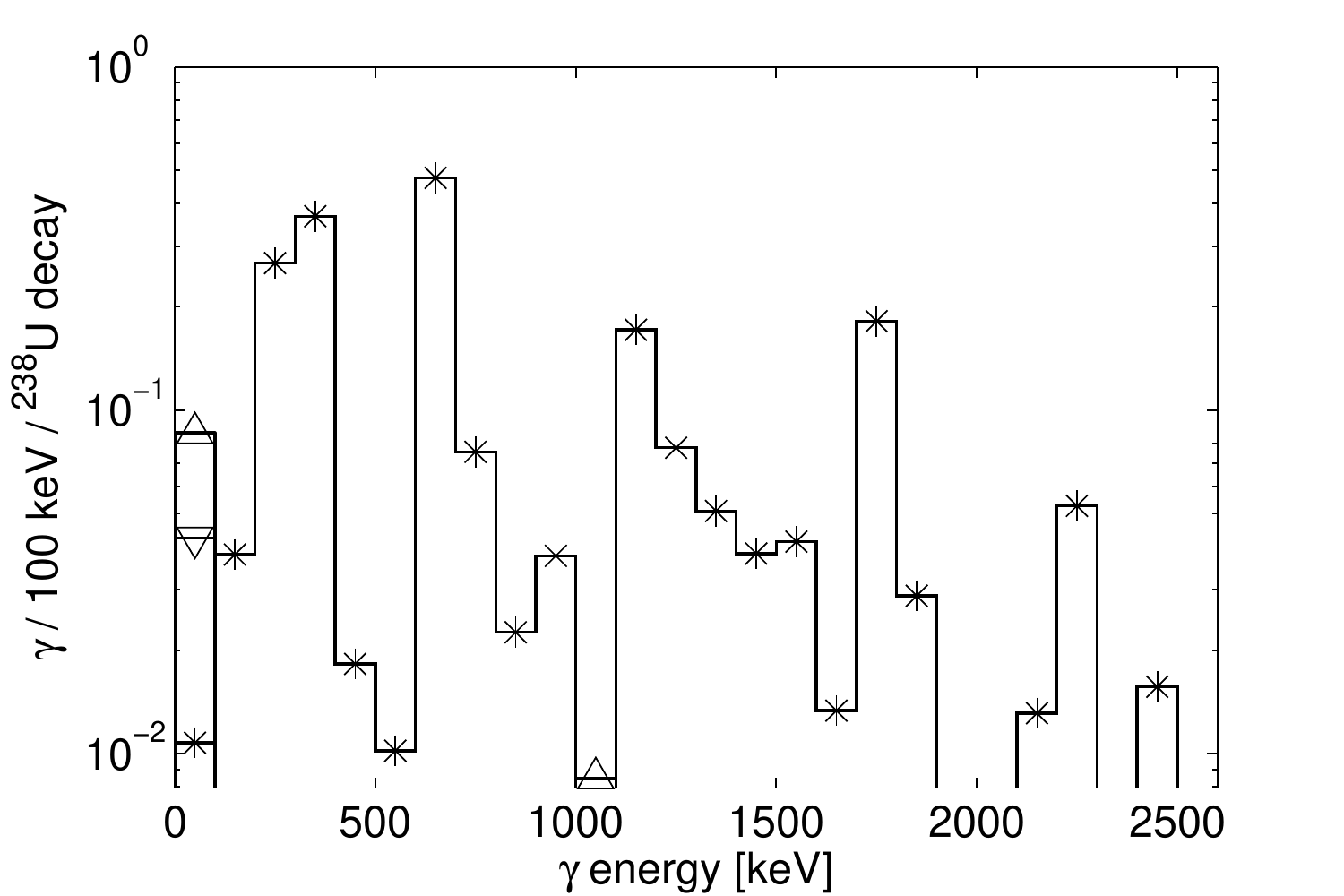}
\par\end{centering}
\caption{\label{Gamma-emission-energies}Gamma energy spectra generated from $^{238}$U decay chain isotopes, assuming full-chain secular equilibrium. Gammas are shown corresponding to the $^{238}$U~early sub-chain (upward arrows), $^{226}$Ra sub-chain (stars), and $^{210}$Pb sub-chain (downward arrows). Data is taken from \cite{nndc}.}
\end{figure}

The $\gamma$ energy emission spectra for the $^{238}$U early sub-chain, $^{226}$Ra sub-chain, and $^{210}$Pb sub-chain are shown in Fig.~\ref{Gamma-emission-energies}. Gamma emission from the $^{238}$U~early sub-chain and $^{210}$Pb sub-chain is subdominant in both intensity and energy to that from the $^{226}$Ra sub-chain. This limits the impact of $^{238}$U chain disequilibrium on ER backgrounds.

ER backgrounds from $^{238}$U disequilibrium is assessed by Monte Carlo, using the detector and emission geometries for one~tonne liquid Ar and liquid Xe detectors described in Sec. \ref{Impact-on-DM-Neutrons}. Gamma energies following the probability distribution shown in Fig.~\ref{Gamma-emission-energies}, for each of the $^{238}$U sub-chains, are thrown from the border between the target material and the water shield, representing activity from detector internals. Position cuts are used to remove multiple-scatter events, as well as events which deposit energy outside of the fiducial volume. Details of these cuts are discussed in Sec. \ref{Impact-on-DM-Neutrons}.

ER event rates are found at low energies and normalized per keV$_{ee}$, where keV$_{ee}$ is defined as energy deposited in the detector by ER. The recoil energy spectrum in both Ar and Xe detectors is flat below the 46.5~keV$_{ee}$ peak from $^{210}$Pb decay. Rates are estimated in the range 1--45~keV$_{ee}$.

\begin{figure}
\begin{centering}
\subfloat[]{\includegraphics[width=0.5\textwidth]{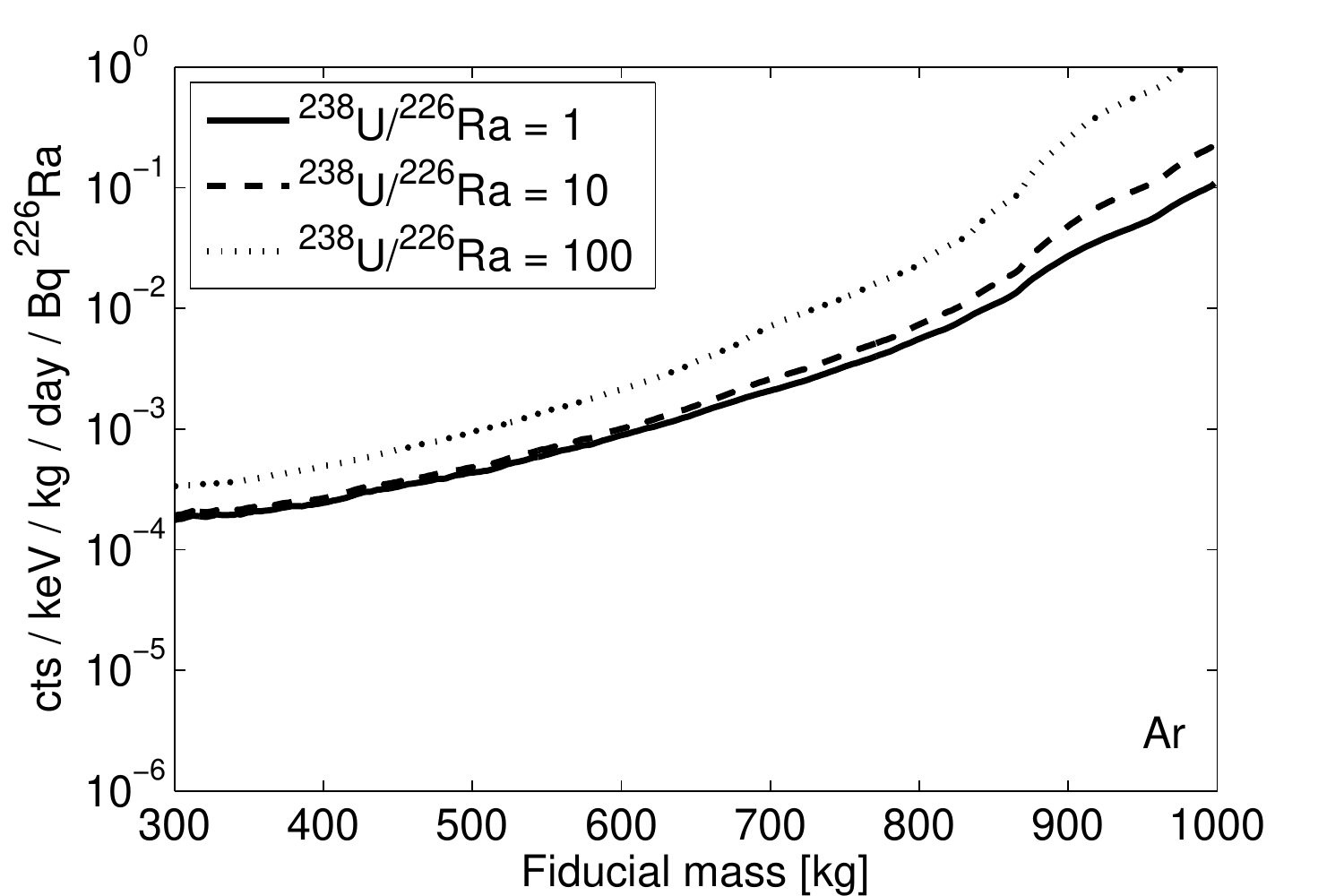}
}
\subfloat[]{\includegraphics[width=0.5\textwidth]{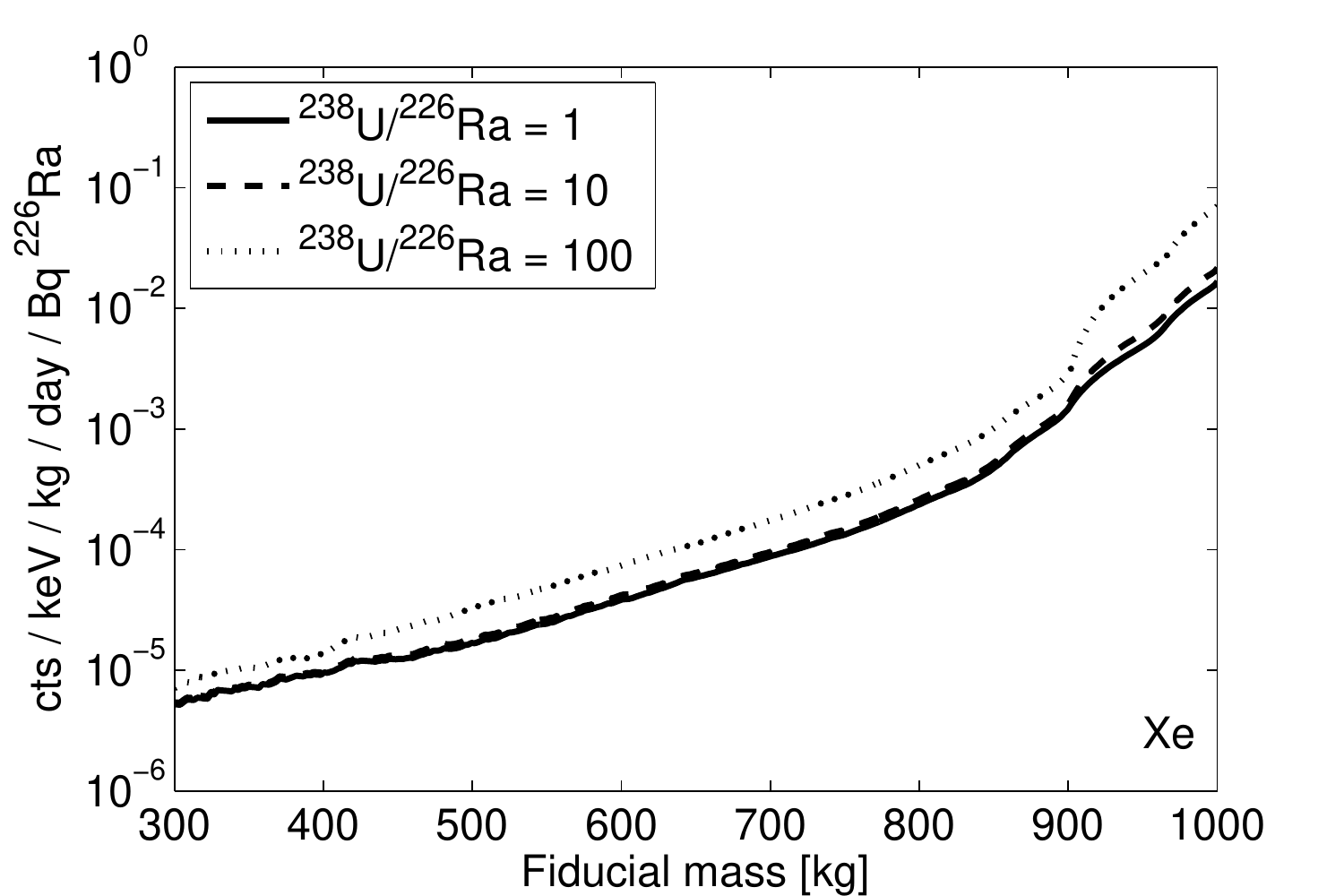}
}
\par\end{centering}
\caption{\label{fid_rate_vs_mass}Fiducial ER rate in simulated one~tonne (a) liquid Ar and (b) liquid Xe detectors from $^{238}$U sub-chain $\gamma$~rays, as a function of fiducial mass. Three scenarios of $^{238}$U~early sub-chain / $^{226}$Ra sub-chain ratio $X$ are shown: $X$=1 (solid), $X$=10 (dashed), and $X$=100 (dotted). The concentrations of $^{226}$Ra sub-chain and $^{210}$Pb sub-chain isotopes are held constant in all scenarios, while the concentration of $^{238}$U~early sub-chain isotopes is varied. Gammas depositing energy in the range 1-45~keV$_{ee}$ are selected, and single-scatter cuts are applied which reject events with energy-weighted standard deviation above 2~cm in radius and 0.5~cm in height.}
\end{figure}

Fiducial rate as a function of mass is shown in Fig.~\ref{fid_rate_vs_mass} for both detectors. Three scenarios are considered, in which the $^{238}$U~early / $^{226}$Ra sub-chain fraction $X$ is 1, 10 or 100. The number of $\gamma$~rays passing the analysis cuts described above and depositing energy within the fiducial region is shown in units of counts~keV$_{ee}^{-1}$~kg$^{-1}$~day$^{-1}$~(Bq~$^{226}$Ra)$^{-1}$. Fiducial volume shape is optimized to yield the lowest activity, while keeping the fiducial boundary convex.

Both Ar and Xe detectors are insensitive to a disequilibrium condition of $X$=10. At 500~kg fiducial mass, fiducial activity is increased by 10\% for the Ar detector, and 9\% for the Xe detector. Event rates are significantly increased for both detectors in the case of $X$=100, with a factor $\times$2.2 increase for the Ar detector, and a factor $\times$2.0 increase for the Xe detector. Holding the event rate constant from the equilibrium case, a factor $\times$100 imbalance results in the loss of 73~kg fiducial mass for the Ar detector and 60~kg for the Xe detector, starting at a 500~kg fiducial in equilibrium. The case of $X$=100 is used as a conservative example; standard $\gamma$ counting techniques are typically sensitive to imbalances a factor $\times$1/2 this number or lower.

\begin{figure}
\begin{centering}
\subfloat[]{\includegraphics[width=0.5\textwidth]{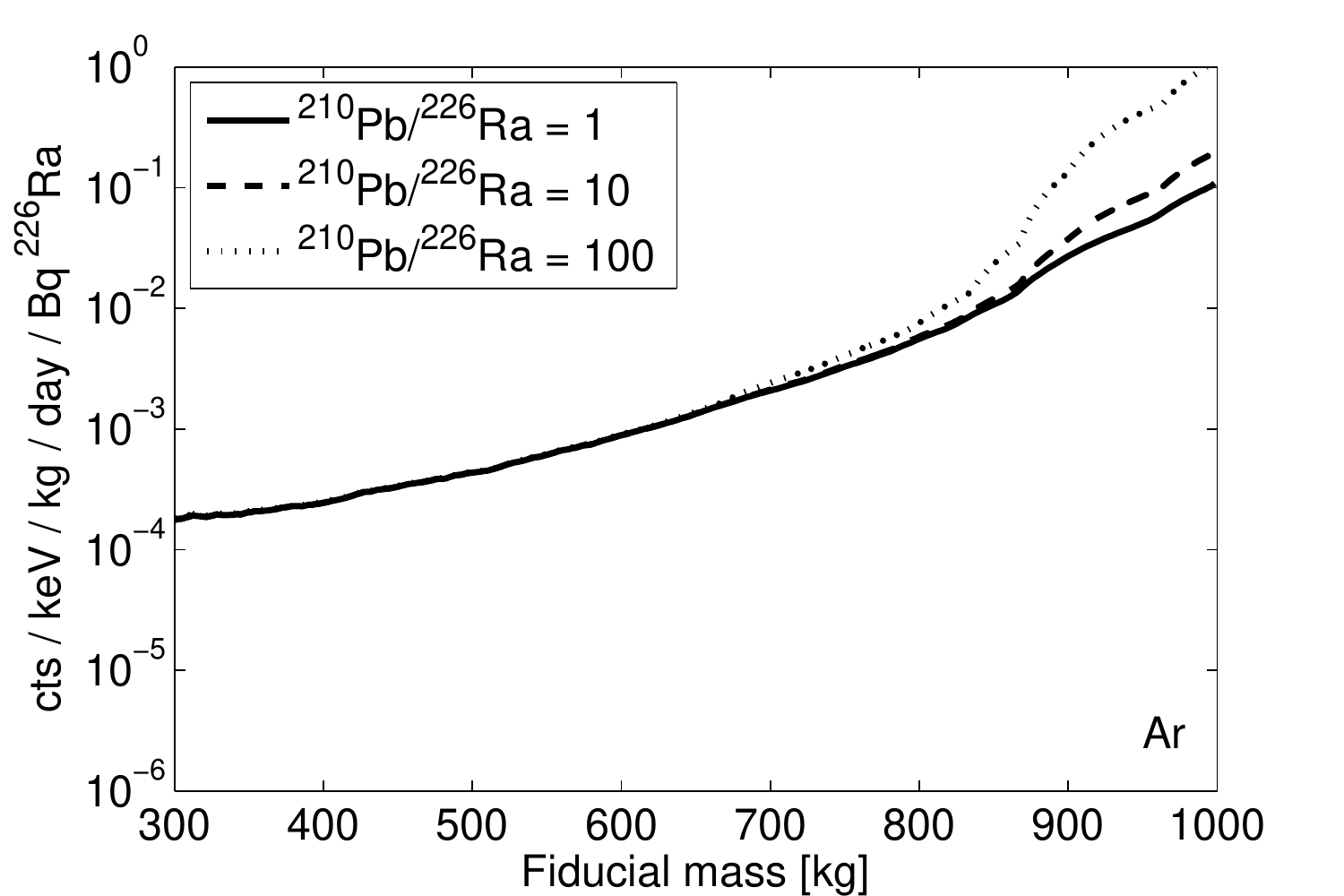}
}
\subfloat[]{\includegraphics[width=0.5\textwidth]{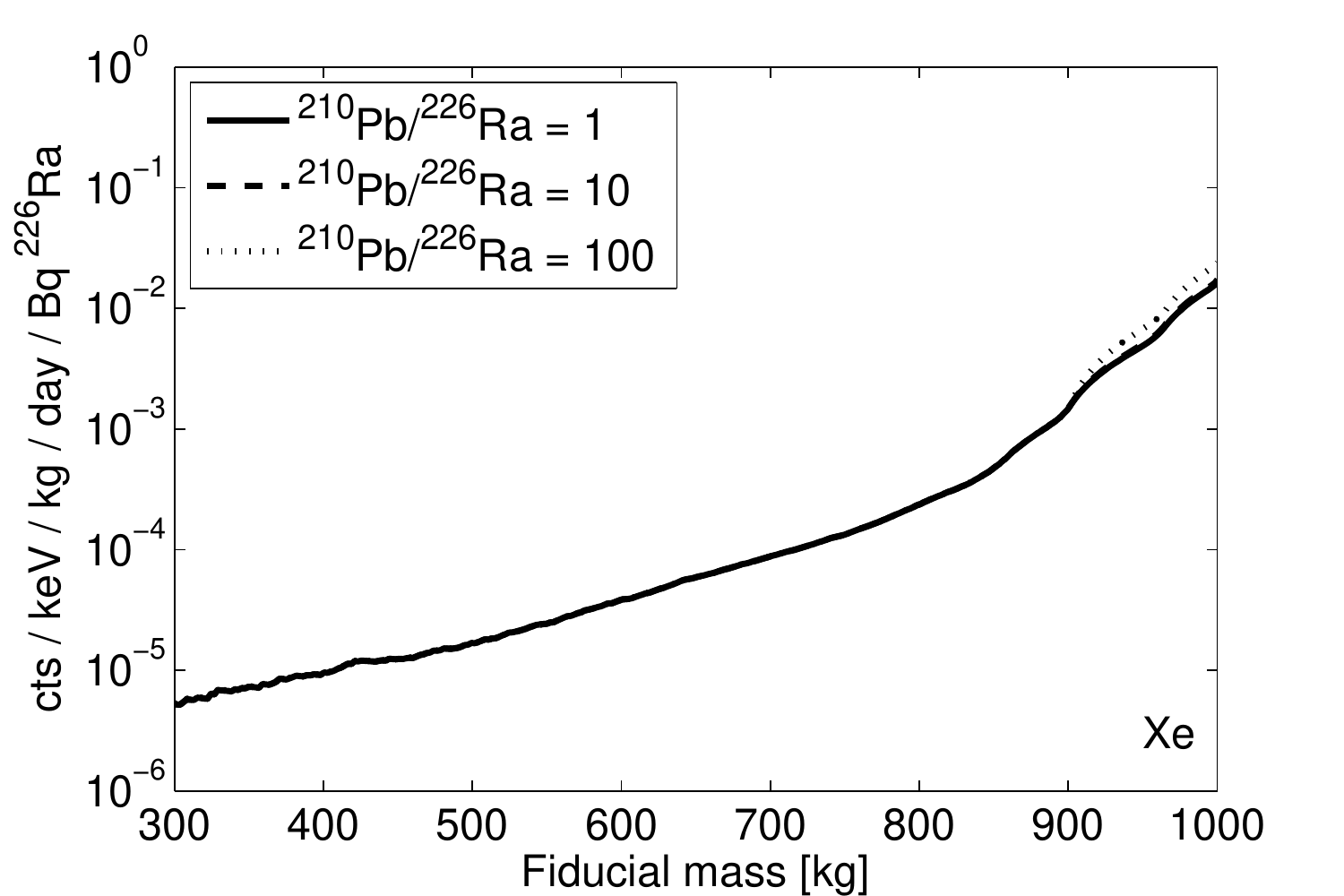}
}
\par\end{centering}
\caption{\label{fid_rate_vs_mass_late}Fiducial ER rate in simulated one~tonne (a) liquid Ar and (b) liquid Xe detectors from $^{238}$U sub-chain $\gamma$~rays, as a function of fiducial mass. Three scenarios of $^{210}$Pb sub-chain to $^{226}$Ra sub-chain rate are shown: $\times$1 (solid, corresponding to $^{238}$U chain equilibrium), $\times$10 (dashed), and $\times$100 (dotted). The concentrations of $^{238}$U~early sub-chain and $^{226}$Ra sub-chain isotopes are held constant in all scenarios, while the concentration of $^{210}$Pb sub-chain isotopes is varied. Cuts are used as described in Fig.~\ref{fid_rate_vs_mass}.}
\end{figure}

For completeness, the impact of an excess of $^{210}$Pb sub-chain rate relative to $^{226}$Ra sub-chain rate is shown in Fig.~\ref{fid_rate_vs_mass_late}. The only significant $\gamma$ emission in the $^{210}$Pb sub-chain comes from the 46.5~keV $\gamma$ from $^{210}$Pb decay. The penetration length of this $\gamma$ is 8~mm in Ar and 0.2~mm in Xe, and thus does not represent a significant background addition after modest fiducial cuts.

\begin{figure}
\begin{centering}
\subfloat[]{\includegraphics[width=0.5\textwidth]{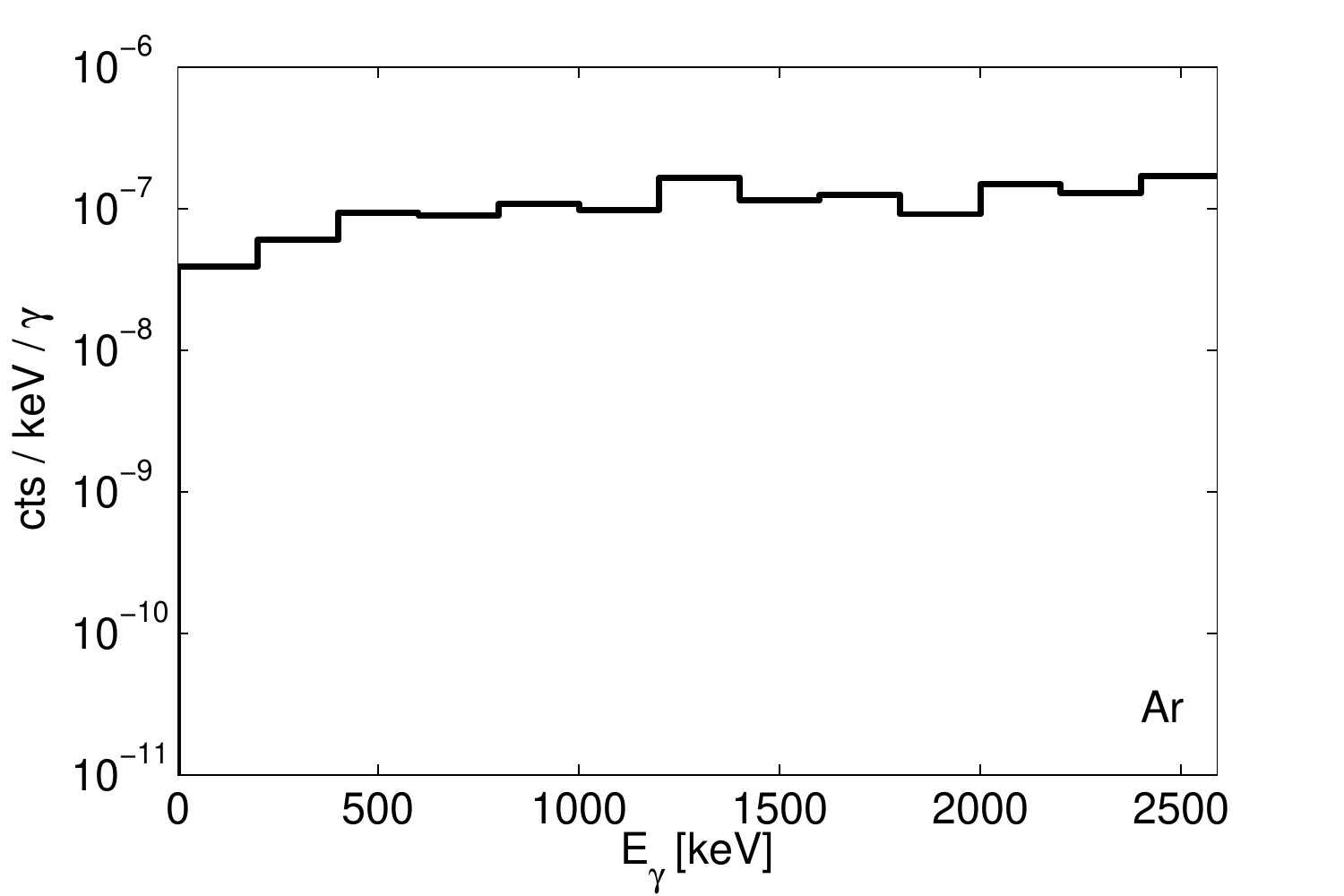}
}
\subfloat[]{\includegraphics[width=0.5\textwidth]{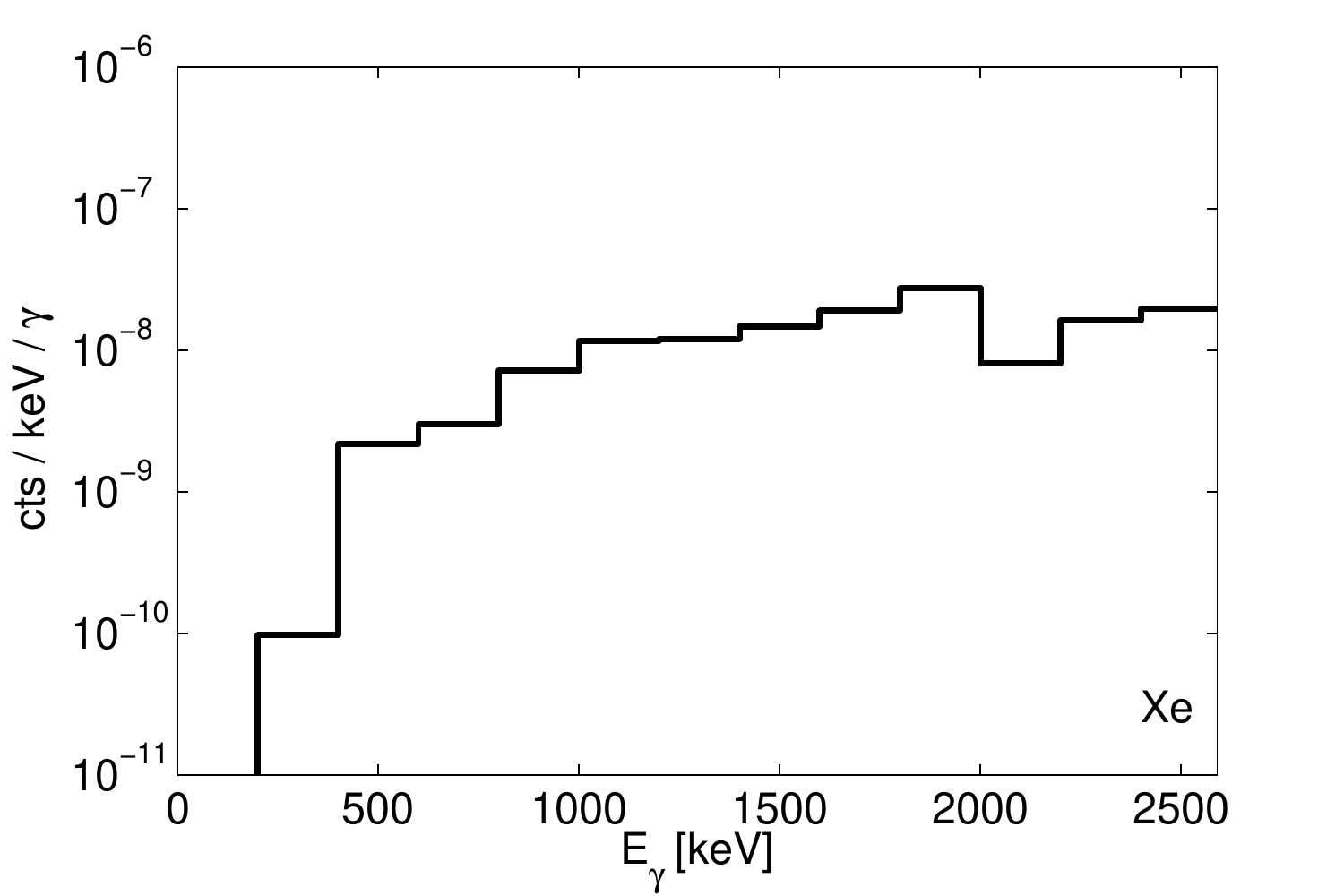}
}
\par\end{centering}
\caption{\label{damage_vs_E}Fraction of $\gamma$~rays generating fiducial backgrounds in simulated one~tonne (a) liquid Ar and (b) liquid Xe detectors, as a function of $\gamma$ energy. Gamma rates are found in a 500~kg cylindrical fiducial mass with 1:1 aspect ratio is used. Single-scatter and energy cuts are applied as described in Fig.~\ref{fid_rate_vs_mass}.}
\end{figure}

The spectrum of $\gamma$~rays passing analysis cuts and contributing fiducial backgrounds is shown in Fig.~\ref{damage_vs_E}. A 500~kg cylindrical fiducial mass with 1:1 aspect ratio was used, with single-scatter and energy cuts used as described above. The photoelectric cross-section is equal to or greater than the Compton cross-section for energies below 75~keV for Ar and 300~keV for Xe. Below these energies, $\gamma$ mean free path drops sharply with energy.

The 500~kg fiducial ER rate in the Ar detector is a factor $\times$30 that of the Xe detector at 500~kg. However, Ar detectors can utilize powerful discrimination against ER events through characterization of the scintillation time constant, potentially reaching ER discrimination power of 10$^{9}$ \cite{ardm,miniclean}. Xenon detectors rely on differences in ionization to scintillation ratio for ER discrimination, with typical values in the range 10$^2$--10$^3$, dependent on electric field strength \cite{xe100,xedisc,zeplin}. In the case of Ar, ER backgrounds from detector components are subdominant to contributions from $^{39}$Ar, a 565~keV endpoint $\beta$ emitter with a rate of 1~Bq~kg$^{-1}$ in natural Ar. For a 500~kg fiducial, backgrounds from detector components only become relevant if $^{39}$Ar is removed to a factor $\times$10$^{-5}$ of its natural value, assuming a 1~Bq concentration of $^{238}$U in equilibrium in detector internals. By using Ar found in natural gas wells, the $^{39}$Ar concentration has been shown to be reduced by a factor $\times$$10^{-2}$ from its typical atmospheric concentration \cite{lowriar}.

\section{Conclusions}

We find that the $^{238}$U~early sub-chain, $^{226}$Ra sub-chain, and $^{210}$Pb sub-chain from the $^{238}$U decay chain have significantly different neutron yields per decay. For typical detector construction materials, the ($\alpha$,n) neutron yield increases by one to three orders of magnitude as $\alpha$ energy increases in the range 4--8~MeV. The $^{238}$U~early sub-chain contributes three $\alpha$ particles per $^{238}$U decay with energies of 4.2--4.8~MeV, at the lowest end of the range. The $^{226}$Ra sub-chain contributes four $\alpha$ particles over a broader range, with energies spanning 4.8--7.8~MeV. If the $^{238}$U~early sub-chain has an order of magnitude greater concentration than the $^{226}$Ra sub-chain, then the total neutron emission rate increases by only $\times$2--4, where the precise increase depends on the target material. A two order of magnitude increase results in a $\times$10 increase in total neutron emission rate. If a material has a low ($\alpha$,n) yield, then neutron emission is dominated by spontaneous fission from $^{238}$U. This leads to a more linear relationship between neutron yield and the ratio of $^{238}$U~early sub-chain to $^{226}$Ra sub-chain. In practice, spontaneous fission events are vetoed by their simultaneous generation of multiple neutrons and MeV $\gamma$ rays, which would interact in the active region of the detector.

Typical $\gamma$ screening with Ge detectors is $\times$20--60 more sensitive to $^{226}$Ra sub-chain decay than to $^{238}$U~early sub-chain decay. This is due to the low emission probability per decay of high-energy gammas ($>$300~keV) from the $^{238}$U~early sub-chain compared to the $^{226}$Ra sub-chain. If $^{238}$U~early sub-chain concentration is $\times$20(60) greater than $^{226}$Ra sub-chain concentration, then neutron emission from high ($\alpha$,n) yield materials such as F and Al is increased by a factor $\times$2(6). Mass spectrometry techniques can potentially detect $^{238}$U directly. However, these techniques are insensitive to $^{238}$U decay chain daughters due to their low concentrations, given their relatively small half-lives. Inhomogeneity in $^{238}$U sub-chain concentrations within material samples can also significantly skew mass spectrometry results, which only sample small amounts of material. Gamma screening techniques sample large masses, and average well over inhomogeneous materials. For large dark~matter detectors, bulk counting is crucial for sampling all detector construction materials, as there can be many tonnes of material used in close proximity to the detector active region.

Background neutron event rates from $^{238}$U~early sub-chain, $^{226}$Ra sub-chain, and $^{210}$Pb sub-chain decays in high ($\alpha$,n) yield materials has been assessed by Monte Carlo for simplified one~tonne liquid Ar and liquid Xe detectors. We find that the Xe detector has a signal-to-background ratio $\times$12 higher than the Ar detector, over neutron events from a similar level of $^{238}$U chain isotope concentration, for identification of 100~GeV WIMPs. The $^{226}$Ra sub-chain dominates neutron backgrounds~per~decay when compared to the $^{238}$U~early sub-chain and $^{210}$Pb sub-chain by an order of magnitude. For a given sub-chain, the Ar and Xe detectors have similar neutron event rates when considering WIMP search energy windows of 50--100~keV$_{nr}$ and 5--30~keV$_{nr}$ respectively, which are typically quoted for these detectors.

Gamma backgrounds for Ar and Xe tonne-scale detectors are impacted negligibly by an excess of $^{238}$U~early sub-chain or $^{210}$Pb sub-chain activity relative to $^{226}$Ra sub-chain activity, for typically quoted $\gamma$ screening limits. For the Xe detector, ER backgrounds from $\gamma$~rays in the 500~kg fiducial are increased by only $\times$1.1 and $\times$2 when $^{238}$U~early sub-chain concentration is increased by a factor $\times$10 and $\times$100 from full-chain equilibrium, respectively. The $^{210}$Pb sub-chain has negligible contribution, even at rates $\times$100 higher than in full-chain equilibrium. The Ar detector has comparable sensitivity to $^{238}$U~early sub-chain and $^{210}$Pb sub-chain excesses as the Xe detector. Argon detector ER backgrounds are typically dominated by $\beta$ backgrounds from $^{39}$Ar, which contribute an ER rate in the 500~kg fiducial which is $\times$$10^{5}$ higher than activity from 1~Bq of $^{238}$U in construction materials.

For experiments in which background rates are dominated by NR events from neutrons, $^{238}$U~early sub-chain and $^{210}$Pb sub-chain excesses can cause a significant increase in experiment backgrounds over expectations based on the assumption of full-chain equilibrium. A $^{238}$U~early sub-chain concentration which is $\times$60 higher than the $^{226}$Ra sub-chain concentration leaves the possibility of a neutron emission rate which is $\times$6 higher than the emission rate from full-chain equilibrium, as discussed above. Experiments with NR-dominated backgrounds would then have a background rate $\times$6 higher than expectations. Experiments which have ER-dominated backgrounds are relatively insensitive to excesses of $^{238}$U~early sub-chain and $^{210}$Pb sub-chain concentrations. ER backgrounds increase by less than a factor $\times$2 for $^{238}$U~early sub-chain excesses within $\gamma$ screening limits, and increase negligibly for a $\times$100 ratio of $^{210}$Pb sub-chain concentration to $^{226}$Ra concentration.

Experiments which have NR-dominated backgrounds can potentially benefit from the use of mass spectrometry screening techniques to directly measure $^{238}$U concentrations in construction materials. These techniques would compliment $\gamma$ radioassay measurements in order to identify potential disequilibrium conditions. Experiments with ER-dominated backgrounds do not gain significant refinement in background expectation from the use of mass spectrometry counting, and are able to rely on screening from $\gamma$ radioassay alone.

\section*{Acknowledgements}

We thank Chamkaur Ghag for his insights into mass spectrometry techniques. This work is supported by the U.S. Department of Energy (DOE) under award numbers DE-FG02-91ER40688 and DE-FG02-13ER42023.

\end{document}